%% file: main.tex
\pdfoutput=1
\documentclass[11pt]{article}
\usepackage{fullpage}
\usepackage{amsmath}
\usepackage{amssymb}
\usepackage{amsthm}
\usepackage{epsfig}
\usepackage{pxfonts}
\usepackage{algorithm}
\usepackage{algorithmic}
\usepackage{color}
\usepackage{natbib}
\usepackage{tikz}
\usepackage{enumerate}
\usepackage{appendix}
\usepackage{soul}

\usetikzlibrary{%
decorations.fractals,%
decorations.shapes,%
decorations.text,%
decorations.pathmorphing,%
decorations.pathreplacing,%
decorations.footprints,%
decorations.markings}

\newcommand{\tm}[1]{\textrm{#1}}
\renewcommand{\P}[1]{\mathbb{P}\left(#1\right)}

\newcommand{\I}[1]{\mathbb{I}\left(#1\right)}

\newcommand{\E}[1]{\mathbb{E}\left[#1\right]}

\newtheorem{theorem}{Theorem}[section]
\newtheorem{lemma}[theorem]{Lemma}

\newtheorem{proposition}[theorem]{Proposition}
\newtheorem{corollary}[theorem]{Corollary}
\newtheorem{conjecture}[theorem]{Conjecture}
\newtheorem{remark}[theorem]{Remark}
\newtheorem{claim}[theorem]{Claim}

\def\t {z}

\usepackage{soul}

\begin{document}


\title{Kidney Exchange in Dynamic Sparse Heterogenous Pools}

\author{Itai Ashlagi  \and Patrick Jaillet \and Vahideh H. Manshadi\thanks{%
Ashlagi: iashlagi@mit.edu. Jaillet:
jaillet@mit.edu.  Manshadi: manshadi@mit.edu. We thank
participants in the NBER workshop on market design  for valuable
suggestions and comments.}}


\date{\today}

\maketitle

\begin{abstract}
Current kidney exchange pools are of moderate size and thin, as they consist of many highly sensitized patients. Creating a thicker pool can be done by waiting for many pairs to arrive. We analyze a simple class of matching algorithms that search periodically for allocations. We find that if only 2-way cycles are conducted, in order to gain a significant amount of matches over the online scenario (matching each time a new incompatible pair joins the pool) the waiting period should be ``very long". If 3-way cycles are also allowed we find regimes {in which waiting for a short period} also increases the number of matches considerably. Finally, a significant increase of matches can be obtained by using even one non-simultaneous chain while still matching in an online fashion. Our theoretical findings and data-driven computational experiments lead to policy recommendations.
\end{abstract}


\section{Introduction}

The need for kidney exchange arises when a healthy person wishes to donate a kidney
but is incompatible with her intended recipient. Two main factors determine compatibility of a donor with a patient: blood-type compatibility and tissue-type compatibility. Two or more incompatible pairs
can form a cyclic exchange so that each patient can receive a kidney from a compatible donor. In addition, an exchange can be initiated by a non-directed donor (an altruistic donor who does not
designate a particular intended patient), and in this case, a chain of
exchanges need not form a closed cycle.

Current  exchange pools are of moderate size and have a dynamic flavor as  pairs enroll over time. Furthermore, they contain many highly sensitized patients (\citet{AshalgiGamarnikRoth}), i.e., patients that are very unlikely to be tissue-type compatible with a blood-type compatible donor. One major decision clearinghouses are facing is how often to search for allocations (a set of disjoint exchanges). On one hand, waiting for more pairs to arrive before finding allocations will increase the number of matched pairs, especially with highly sensitized  patients, and on the other hand, waiting is costly.
This paper studies this intrinsic tradeoff between the waiting time before searching for an allocation,  and the number of pairs matched under myopic, ``current-like", matching algorithms.

Today, clearinghouses for kidney exchange adopt matching algorithms that generally search for  allocations with the maximum number of matches in the existing pool up to some tie-breaking rules.\footnote{Ties are broken mostly in favor of highly sensitized pairs.} We analyze a similar algorithm,  hereafter called {\it Chunk Matching} ($CM$), which accumulates a given number of  incompatible pairs, or a chunk, before searching for an allocation in the pool, and perform sensitivity analysis on the chunk size. Besides answering this design question, through our analysis, we indicate the significant role of the sparsity level of the underlying compatibility graph. We show that if each patient  has enough compatible donors in the pool, even making immediate irrevocable allocations is almost optimal (this is consistent with~\citet{Utku}, who analyzed the optimal mechanism when there is no tissue-type incompatibilities).  However, in practice (even in the horizon of a couple of years) pools are of moderate size, containing many highly sensitized pairs.

\citet{RothKidneyQJE} first proposed a way to organize kidney exchange integrating cycles and chains. Logistical constraints required that  cycles will involve no more than 2 patient-donor pairs (\citet{RothKidneyJET,RothKidneyAERPP}).\footnote{Cyclic exchanges need to be conducted simultaneously since it is required that a donor does not donate her kidney before her associated patient receives a kidney.} Subsequent work suggested that a modest expand of infrastructure, that is allowing  only slightly larger, 3- and 4-way
exchanges would be efficient (\citet{RothKidneyAER}, \citet{AshlagiRothIR})  in large static pools.\footnote{Today,  kidney exchange is practiced by a growing number of hospitals and formal and
informal consortia (see~\citet{RothHahn}). \citet{ABS07} have proposed an algorithm that
works in practice for finding cycles in relatively large size exchange pools.} \citet{Utku} has initiated the study of dynamic kidney exchange showing a closely related result to the static case. These studies assume either implicitly or explicitly that no tissue-type incompatibilities exists.


{However,} as data reveals, most patients in exchange pools are either very hard  or very easy to match (high and low sensitized) {and indeed, a large fraction of them are very highly sensitized}. Here, we focus on these two types, high and low sensitized, while abstracting away from blood-type compatibility.\footnote{Equivalently we assume that all  pairs in the pool are blood-type compatible.} We consider a discrete time model with $n$ pairs that arrive sequentially, one pair at each time period. Each arriving pair is sampled from a bi-modal distribution independently. This model is a dynamic version of \citet{AshalgiGamarnikRoth}. {In the static case, such a model proves successful in capturing the structure of the current exchange pools, and explaining the effectiveness of long chains that are widely used in practice (\citet{AshalgiGamarnikRoth}).} One way to think of $n$ is the number of pairs in the ``relevant" horizon, considered to be the longest reasonable period of waiting.

We first study  the performance of the $CM$  algorithm when it searches for allocations limited to cycles of length 2. In our first main result we show that if the waiting period between two subsequent match runs is  a sublinear  function of $n$, the $CM$ algorithm matches approximately  the same number of pairs as the online scenario (i.e., when searching for a maximum allocation every period without waiting) does (Corollary \ref{cor:sublinear_vs_online}).
Waiting, however, a linear fraction between every two runs, will result in matching linearly more pairs compared to the online scenario (Theorem \ref{thm:linear}).
We generalize our results under more flexible waiting periods, allowing for easy and hard to match pairs wait a different amount of time,


We then analyze $CM$ when cycles of  length both 2 and 3 are allowed. We show that for some regimes, sub-linear waiting (even with only easy to match nodes)  will result in a linear addition of matches comparing to the online scenario
(Theorem \ref{prop:sublinear:3way:LWaiting}).


As chains have become very effective, it is important to study their benefit and analyze the efficiency of matching with chains (chains are initiated by a non-directed donor). A major  difficulty with chains is that they can be of arbitrary length.\footnote{Note that a chain can be conducted non-simultaneously while keeping the restriction that every patient receives a kidney before her associated donor gives one.}
We show that in the online scenario  adding one non-directed donor will increase linearly the number of matches that the $CM$ algorithm will find over the number of matches it will find without a chain (see Theorem \ref{thm:chain}). This  can be viewed as the ``online version" of  the  result by  \citet{AshlagiAlgChains}, who show that in a static  large sparse pool  allowing a single long chain will increase linearly  the number of matched pairs.


In  all our results in which waiting proves to be effective, the additional matches correspond to pairs with highly sensitized patients.  Pairs with low sensitized patients will (almost) all be matched regardless of the size of the chunk in each match run. These findings explain computational simulations using clinical data (Figures \ref{figCMsimulationTwoWay} and \ref{figCMsimulationThreeWay}).

Our results are given for a pool in which the highly sensitized pairs have on average a constant number of compatible donors. We extend the results for pools with increasing density levels.  The results show that the more ``dense" the pool, the less the clearinghouse should wait in order to match linearly more pairs than the online solution. This set of results again indicate the crucial effect of sparsity of the compatibility graph  and thus the importance of accurate modeling of it.

Our results may be of independent interest to the literature on dynamic matching in random graphs. Kidney exchange serves well as an example for which we have distributional information on the underlying graphs, thus we can exploit this information to make analysis and prediction far more accurate than the worst-case analysis can do. We believe our average-case analysis can have implications beyond the kidney exchange and can be applied to other dynamic allocation problems with such distributional information.


While this paper focuses on  kidney exchange, there are  many dynamic markets for barter exchange for which our findings apply.
There is a growing number of  websites that accommodate a marketplace for exchange of goods (often more than $2$ goods), e.g. ReadItSwapIt.com  and Swap.com. In these markets, the demand for goods, cycle lengths and waiting times play a significant role in efficiency.

\subsection{Related work}

The literature on dynamic kidney exchange is in its very beginning. \citet{Utku} initiated dynamic kidneys exchange focusing on large dense pools. Our work deviates from that model significantly not only by analyzing sparse pools, but also by abstracting away from the blood types and focusing on the tissue-type compatibility. Further, our approach to study dynamic kidney exchange is combinatorial and is based on the structure of the underlying random graph while \citet{Utku} takes a dynamic programming approach.

\citet{ProcacciaSandholm} conduct computational simulations in the dynamic settings to understand the benefit of chains. \citet{ProcacciaSandholm2} study dynamic optimization and propose an algorithm that assigns weights to different matches using future stochastic sampling. These studies both use dense compatibility graphs.

The problem of online matching (equivalent to our online scenario with only two-ways) arises naturally in information technology applications such as online advertising in which advertisements need to be assigned instantly to queries searched or webpages viewed by  users. The study of online matching was initiated by  \citet{kvv}, in which they analyze the problem in  adversarial settings with no probabilistic information about the graph. Several follow up papers, studied the problem in settings that limit the power of the adversary. \citet{aryanak_randominput} studied the model in which the underlying graph has unknown distribution. \citet{aryanak_stmatching} noticed that in applications such as online advertising there is information about the graph structure, and they analyzed a  model where the graph distribution belongs to a certain class. \citet{mos} studied the same problem with a general known distribution. Note that here we focus on one special class of distributions; however, unlike the  computer science literature, we consider various regimes of waiting (and not just the online scenario).

\citet{Mendelson} analyzed the behavior of a clearinghouse in a dynamic market with prices in which sellers and buyers arrive over time according to a given stochastic process. Similar to our work, he considers a mechanism in which the clearing prices are computed periodically, and he studies the market behavior for different time (period) scales.

\input{Model}
\input{chunk}
\input{GeneralSparseGraphs}
\input{chainPcycle}

\input{extension}

\section{Discussion}
\label{sec:discussion}

Previous theory for kidney exchange dealt with dense graphs, finding that efficiency can be obtained via short cycles.
In  dense graphs,  waiting in order to accumulate incompatible pairs is also not an issue.
Recently it was shown that the pools we observe in practice are very sparse with many highly sensitized patients for which the previous theory does not hold.
This raises the question of the tradeoff between waiting for more pairs to arrive and the number of matches one obtains.
We initiate here this direction, studying a class of algorithms that find a maximum allocation after every $x$ pairs arrive.

We find that in sparse graphs, when only short cycles are allowed, it is only when the algorithm waits for significant amount of pairs to arrive
that it will match significantly many more pairs than the online scenario does.

It has been shown that even a single unbounded chain, beginning with a non-directed donor,  increases efficiency significantly in large static sparse pools beyond just short cycles.
We show here the dynamic version of this result for the online setting: we find that in the online scenario with a single non-directed donor the algorithm will match linearly many more pairs than without the non-directed (this result assumes that either, in both settings, cycles can be of length at most 2, or the pool only contains pairs with hard to match patients). We conjecture that the last result holds also when cycles of length $k>2$ are allowed.

Our results suggest that if a centralized clearinghouse cannot afford to wait ``too long", then online matching is a good solution.\footnote{No multi-hospital kidney exchange program in the US is currently waiting more than a month before finding allocations.} Dynamic chains can be used to reduce the disadvantage of online matching over matching with waiting.

Our work leaves many more questions than answers.  While waiting times are part of the matching algorithms, we do not study the average waiting time of pairs and only focus on the number of allocations. Observe that there is very tight correspondence between the number of matches and waiting time. Thus, although with linear size chunks one obtains more matches,  the average waiting time may increase. Related to this, it will be interesting to study the steady state of the system. Another direction is whether non-myopic algorithms can improve both the waiting times and the number of pairs matched.
As pairs wait to be matched, designing mechanisms that take into account incentives for  patients (e.g., \citet{RothKidneyJET}) and hospitals  (e.g., \citet{AshlagiRothIR}) becomes an intriguing task.

Thickness is an important property for efficiency in  market design. Kidney exchange clearinghouses can create a thick market at the cost of  waiting for many pairs to arrive. Tradeoffs between unraveling (waiting before entering the market) and thickness are of practical importance in many other markets such as job markets and markets for graduate students (see e.g., \citet{MRothThick}). Our theory can serve as a building block for studying such tradeoffs and for the study of  implementing ``efficient" outcomes in the long run when agents have preferences.


%
%
%
%

\bibliographystyle{plainnat}
\bibliography{kidneysbib}

\input{appendix}

\end{document}

%% file: Model.tex
\section{Dynamic compatibility graphs and empirical findings}
\label{sec:model}

In a kidney exchange pool there are patients with kidney failure, each associated with an incompatible living donor, and  non-directed donors (NDDs).\footnote{Pairs that are compatible would presently go directly to transplantation and
not join the exchange pool although \citet{RothKidneyAERPP} and %
\citet{AltruisticUtku}  study the advantage of adding such pairs to the pool.}
The set of incompatible pairs and NDDs  in  the pool, $V$, induces a {\bf compatibility  graph}  where  a directed arc from $v_1$ to $v_2$ exists if and only if the donor of pair $v_1$ is compatible with the patient of pair $v_2$.\footnote{In practice a minority of patients enroll with multiple donors. One can extend the model appropriately to capture this multiplicity.}

A k-way cycle is a directed cycle in the graph involving $k$ pairs. A chain is a  directed path starting from an NDD. A $k$-{\it allocation} or a $k$-{\it matching} is a set of disjoint cycles each  of size at most $k$. In practice,  cycles of size at  most 3 are considered due to incentive and logistic reasons.

In a {\em dynamic compatibility graph} the pairs (nodes) arrive sequentially one at a time, and at each time step a centralized program can decide on an allocation and remove the participating nodes in that allocation from the graph.   In this paper, we  analyze an algorithm  that finds a maximum allocation every given number of periods  (the algorithm is described in detail in Section \ref{sec:chunk}). Such an algorithm is used in practice and a question faced by centralized programs is how often to search for an allocation.
In the next section, we discuss some empirical findings that will motivate our modeling assumptions.

\subsection{Empirical findings}
\label{sec:empirical}

As opposed to earlier studies that focused on blood types and ignored market size and sensitization of patients, \cite{AshalgiGamarnikRoth} have shown, using historical data from the Alliance for Paired Donation (APD), that  sensitization of patients plays a crucial role in efficiency (they were interested in a maximum allocation in a static pool).
Each patient has a level of percentage reactive antibodies (PRA) that captures how likely a patient will not match a random blood-type compatible donor in the population. The lower the PRA the more likely the patient will match a random donor. \cite{AshalgiGamarnikRoth} find that the percentage of high PRA (PRA above 80\%) in the pool is significantly higher than what previous studies have assumed to support earlier theoretical  findings (see also \citet{SaidmanTrans} and \citet{RothKidneyAER} for such simulations).
They further find that  among patients that have high PRA the average PRA is above 95.

\begin{figure}[h]
\centering
{\includegraphics[width=0.8\textwidth,natwidth=350,natheight=500]{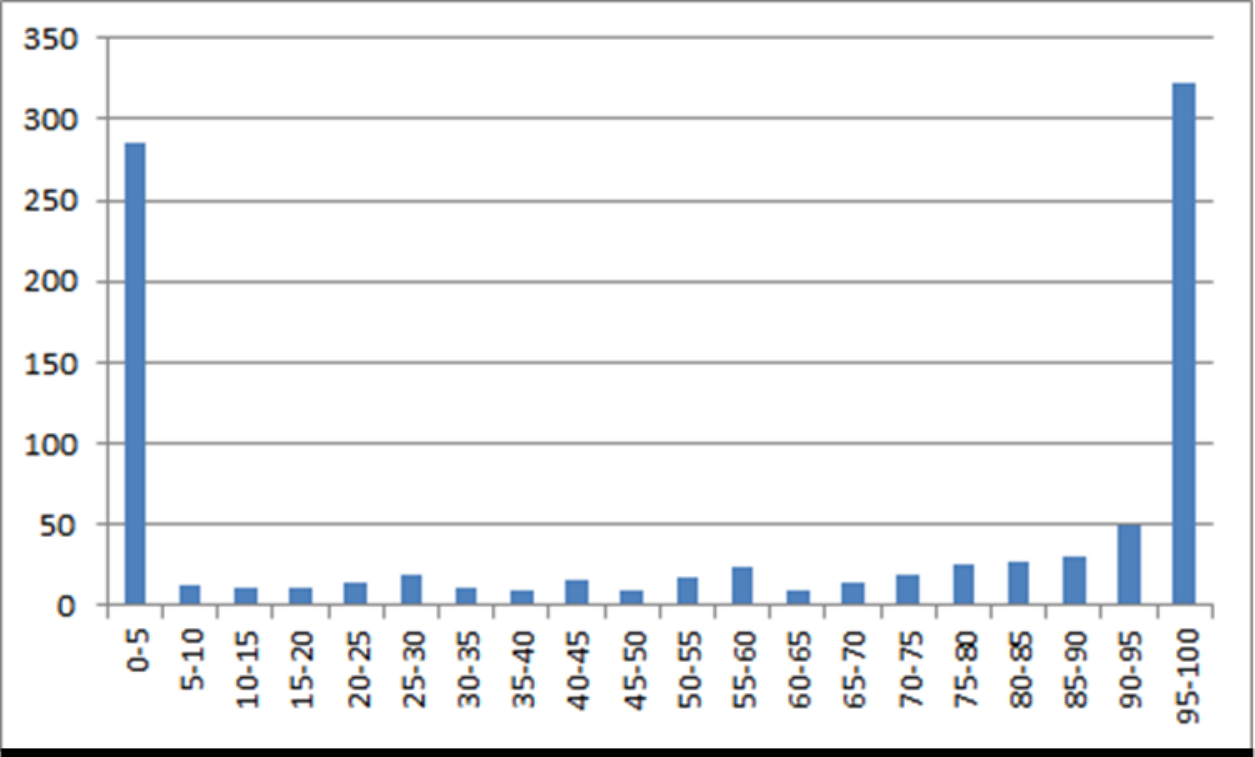}}
\caption{{\protect \footnotesize {PRA distribution of the patients in the exchange pool.}}}
\label{fig:praDist}
\end{figure}

Figure  \ref{fig:praDist} provides a distribution of PRA  in the historical exchange pool of the APD. Note that most pairs have either very high PRA (above 95) or relatively low PRA.

\begin{figure}[tbh]
\centering
{\includegraphics[width=0.8\textwidth,natwidth=350,natheight=500]{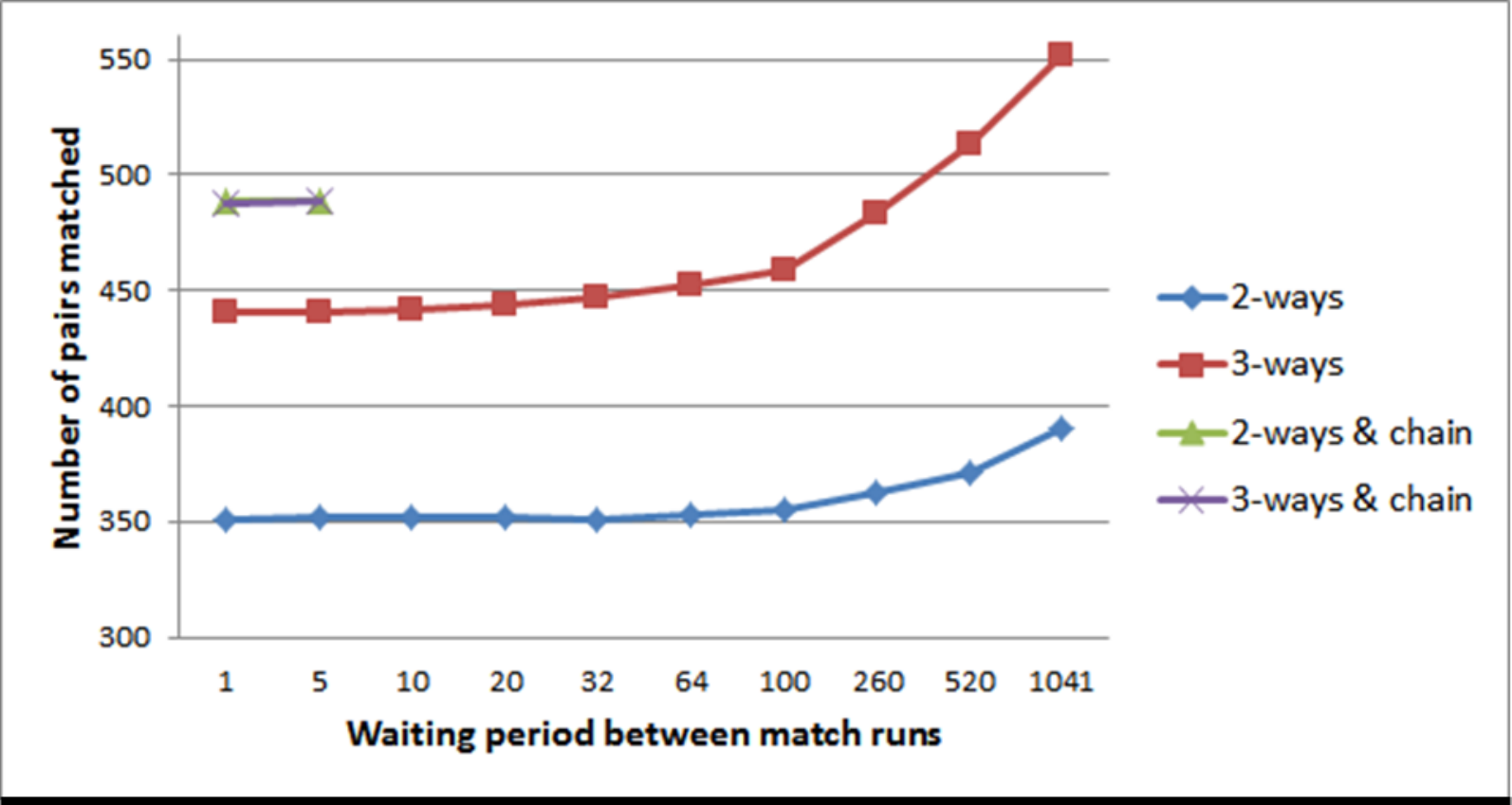}}
\caption{{\protect \footnotesize {Number of matched pairs vs. waiting for $x$ new patients to arrive. Scenarios are: (i) diamond points: only 2-way cycles are allowed; (ii) square points: 2 and 3-way cycles are allowed; (iii) triangle points: 2-way cycles and a single chain (the non-directed donor arrives at the first period); and (iv) cross points: 2 and 3-way cycles and a single chain. Note that triangle and cross points overlap.}}}
\label{figCMsimulationTwoWay}
\end{figure}

\begin{figure}[tbh]
\centering
{\includegraphics[width=0.8\textwidth,natwidth=350,natheight=500]{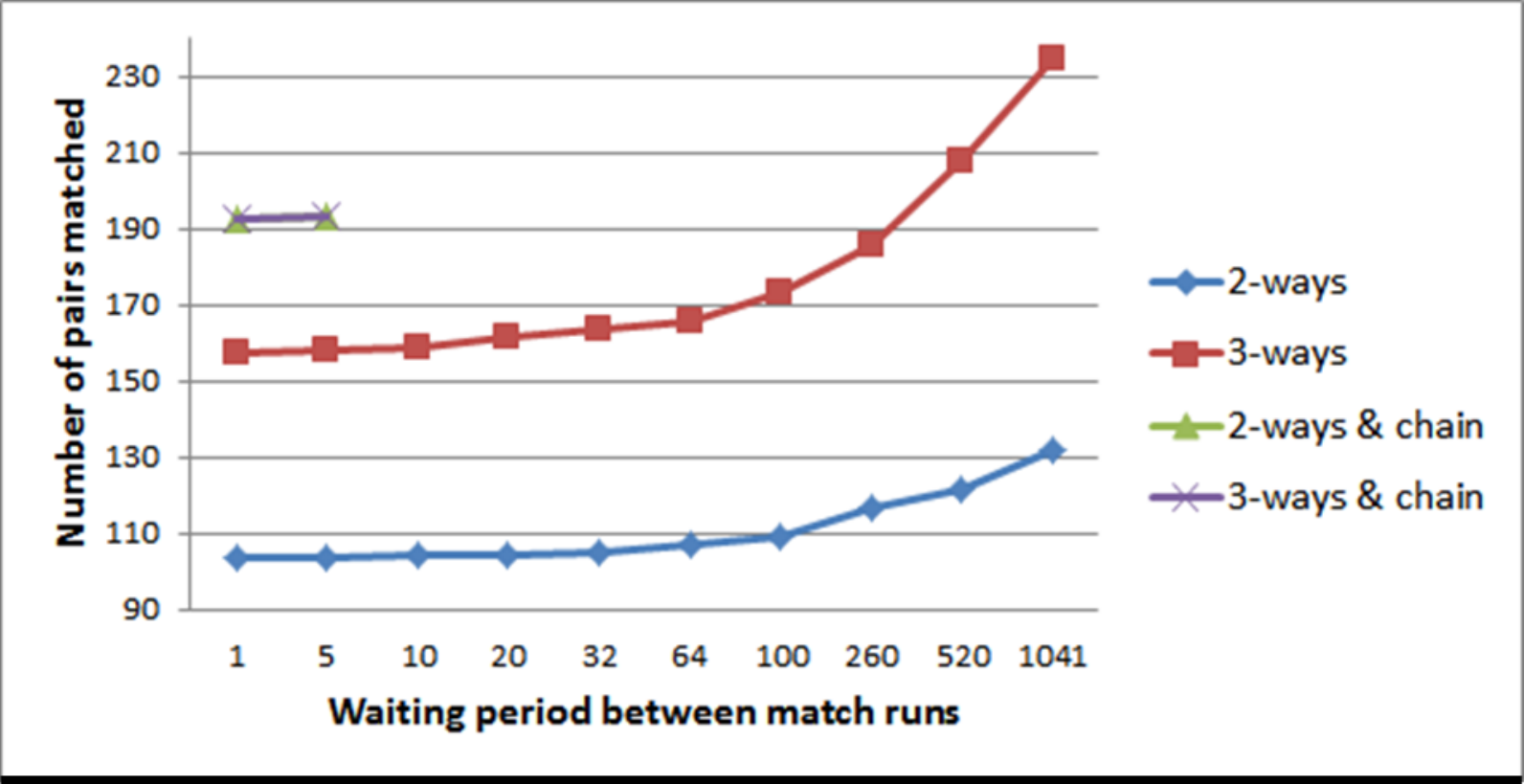}}
\caption{{\protect \footnotesize {Number of highly sensitized patients matched  vs. waiting for $x$ new patients to arrive. Scenarios are: (i) diamond points: only 2-way cycles are allowed; (ii) square points: 2 and 3-way cycles are allowed; (iii) triangle points: 2-way cycles and a single chain (the non-directed donor arrives at the first period); and (iv) cross points:  2 and 3-way cycles and a single chain. Note that triangle and cross points overlap.}}}
\label{figCMsimulationThreeWay}
\end{figure}

Next we show some initial empirical results when matching over time. We have conducted computational experiments in which we use clinical data from  over two years. For each donor and patient  we can determine using their medical characteristics (blood type, antibodies, antigens) whether they are compatible even if they  have not been present in the pool at the same time in practice. We test how many matches are obtained when we search for an allocation after every $x$ pairs have arrived. For each scenario we conduct 200 trials, in which we permute the order in which the pairs arrive. Figure \ref{figCMsimulationTwoWay} plots the number of pairs matched under different waiting periods. Scenarios differ by the length of cycles that are allowed (2-ways, or up to 3-ways, both with or without a single non-simultaneous chain). Figure \ref{figCMsimulationThreeWay} is similar only counting the number of highly sensitized patients that were matched.


{First note that when matching in an online fashion, allowing cycles of length $3$ is quite effective, and increases the number of matches significantly compared to the case when we only allow 2-way exchanges. Further, if we add even a single non-simultaneous chain (by adding one altruistic donor at the beginning), we will match many more pairs.}

Interestingly, {in 2-way matching}, a significant increase in the number of matches occurs only when the waiting period is ``large".
{When allowing 3-ways, short waiting does result in slight improvement, but again a significant gain is only achieved after waiting for a long period.}
{In this paper, we ground the theoretical foundations that explain these behaviors in dynamic matchings in ``sparse'' graphs.}



Sections \ref{sec:chunk}-\ref{subsec:proof:linear} focus on allocations with 2-way cycles and we extend the theory for chains and 3-way cycles in Section \ref{sec:ChainPCycle}.
In the next section we provide our modeling assumptions.

\subsection{A dynamic random compatibility graph}
\label{subsec:model}

\begin{figure}[htbp]
 \begin{centering}
  \input{BasicModel.tex}
  \end{centering}
  \caption{Arc formation in the dynamic kidney exchange; An H-L $2$-way cycle (or, alternatively, an undirected edge); An H-H-L $3$-way cycle.}
  \label{fig:edge_prob}
\end{figure}
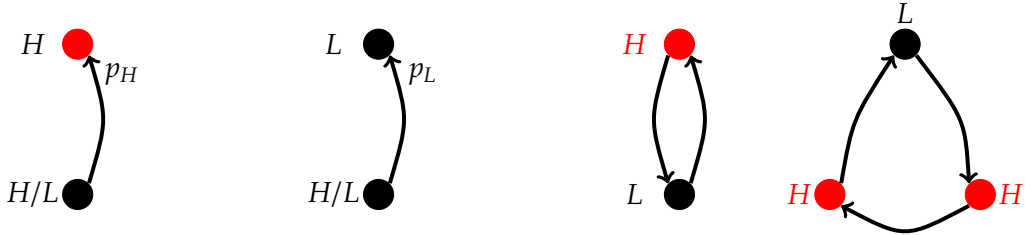

In a {\it dynamic kidney exchange graph}, there are  $n$ patient/donor incompatible pairs which arrive sequentially at times $t = 1, 2, \ldots, n$.\footnote{Without loss of generality, assume $n$ is a power of $2$.} Each pair corresponds to a node in the graph. Each node is one of two types, L (low PRA) or H (high PRA) capturing whether the patient of that node is easy or hard to match. The probability that a node is of type  H is given by $0\leq \rho \leq 1$.   When joining the pool, the arriving node $i$ forms directed edges to the existing nodes. If node $i$ is of type H (L), it forms an incoming arc with any of the existing nodes independently with probability $p_H$ ($p_L$).  Further, it forms outgoing arcs to each L node (H node) independently with probability $p_L$ ($p_H$) (See Figure \ref{fig:edge_prob}).
At each time step, there is an underlying compatibility graph and  the centralized program can find an allocation and remove the participating nodes in that allocation from the graph. We mostly restrict attention to $k=2$, and extend our results to $k=3$ in Section \ref{subsec:3Way}. For the case of $k=2$ it will be convenient  to  reduce  cycles of length two to undirected edges and remove the rest of the directed arcs from the graph. Allocations with $k=2$ are just matchings in the reduced graph.

Note that the maximum number of matches can be obtained after waiting for all pairs to arrive. This is called the {\em offline} solution. As we shall see, the performance of any dynamic allocation scheme depends heavily on the {\em sparsity level} of the compatibility graph which is determined by the arc probabilities $p_H$ and $p_L$. We will assume that $p_L = p$, i.e., an L-patient can receive a kidney from any donor with a fixed probability that is independent of the pool size. On the other hand,  H-patients are much harder to match, and the historical data suggests their in-degree is very ``small" relative to the pool size (see Section \ref{sec:empirical}).

One can in fact show that when $n$ grows large, even in a pool of only highly sensitized pairs, if $p_H$ were to be chosen as a constant, then {an online greedy algorithm would match almost all of the pairs}:

\begin{lemma}
\label{lem:greedy_dense}
Suppose $\rho=1$, i.e., all the arriving nodes are H nodes and let $p_H$ be a constant. An online greedy algorithm, which finds a maximum number of matches after each node's arrival, will match in expectation a total of $n - o(n)$ nodes over the $n$ arrivals.
\end{lemma}
Lemma \ref{lem:greedy_dense} is proven in Appendix \ref{appendixA}. It is  related to the result by \citet{Utku} which assumes no tissue-type incompatibilities; in  both cases the graphs are ``dense" enough, or alternatively large enough, so there is no need to wait before matching.  This contradicts our findings that waiting longer will result in considerably  more matches. If the pool size becomes significantly larger in the future (alternatively, arrival rates become larger), then such results become more relevant.

In addition, random graph results imply that in a large dense graph all blood-type compatible pairs can be matched to each other using only 2-way cycles. As has been seen in \citet{AshalgiGamarnikRoth}, this is not the case (see for example Figure 2 in their paper).
Thus to capture the sparsity we observe in practice, we let $p_H = c/n$, where $c>0$. In Section \ref{sec:dense}, we generalize our results for other sparsity levels $p_H = cn^{-1 + \sigma}$ for any $0 \leq \sigma < 1$. 

\vspace*{0.5cm}
\noindent{\it Remark:} In practice $p_H$ should  only depend on medical  characteristics of the patient regardless of the population size. Setting $p_H$ to be a small number may seem to be a reasonable assumption. However, in our ``relevant" horizon we observe only a small number of pairs to arrive, approximately $\Theta({1/p_H})$, which brings us to the proposed model, linking $n$ and $\Theta({1/p_H})$. For further discussion see \citet{AshalgiGamarnikRoth} from which we adopt these probabilistic assumptions.  Our model also abstracts away from blood type compatibilities and focuses on the sensitivity of patients as the sensitivity of patients is of first order importance in maximizing the number of matches in sparse pools.


%% file: BasicModel.tex
%
%
%
%

\def\radius{2.6}
\def \Pointsize {1.4pt}
\begin{tikzpicture}[pre/.style={<-,shorten <=1.5pt,>=stealth,thick}, post/.style={->,shorten >=1pt,>=stealth,thick}]
\tikzstyle{every node}=[draw,shape=rectangle,minimum size=5mm, inner sep=0];
\tikzstyle{edge} = [draw,thick,-]
\tikzstyle{every node}=[shape=circle,minimum size=8mm, inner sep=0];

\draw [fill, red](0,2) circle [radius=0.2];
\draw [fill](0,0) circle [radius=0.2];

\node [right] at (-1,2) {$H$};
\node [right] at (-1,0) {$H/L$};
\node [left] at (1,1.6) {$p_H$};

\draw[line width=1.5pt] [->] (0.15, 0.15) .. controls(0.4,1) .. (0.15, 1.85);

\draw [fill](0+4,2) circle [radius=0.2];
\draw [fill](0+4,0) circle [radius=0.2];

\node [right] at (-1+4,2) {$L$};
\node [right] at (-1+4,0) {$H/L$};
\node [left] at (1+4,1.6) {$p_L$};

\draw[line width=1.5pt] [->] (0.15+4, 0.15) .. controls(0.4+4,1) .. (0.15+4, 1.85);

\draw [fill,red](0+8,2) circle [radius=0.2];
\draw [fill](0+8,0) circle [radius=0.2];

\node [right,red] at (-1+8,2) {$H$};
\node [right] at (-1+8,0) {$L$};

\draw[line width=1.5pt] [->] (0.15+8, 0.15) .. controls(0.4+8,1) .. (0.15+8, 1.85);
\draw[line width=1.5pt] [->] (-0.15+8, 1.85) .. controls(-0.4+8,1) .. (-0.15+8,0.15);

\draw [fill,red](0+10,0) circle [radius=0.2];
\draw [fill,red](0+12,0) circle [radius=0.2];
\draw [fill](0+11,2) circle [radius=0.2];

\node [left,red] at (0+10,0) {$H$};
\node [right,red] at (0+12,0) {$H$};
\node [above] at (0+11,2) {$L$};


\draw[line width=1.5pt] [->] (-0.15+12, -0.15) .. controls(11,-0.6) .. (0.15+10, -0.15);
\draw[line width=1.5pt] [->] (0.15+10, 0.15) .. controls(10.3,1) .. (-0.15+11,-0.15+2);
\draw[line width=1.5pt] [->] (0.15+11, 1.85) .. controls(12-0.2,1) .. (-0.15+12, 0.15);

\end{tikzpicture}

%% file: chunk.tex
\section{Chunk matching - main results}
\label{sec:chunk}

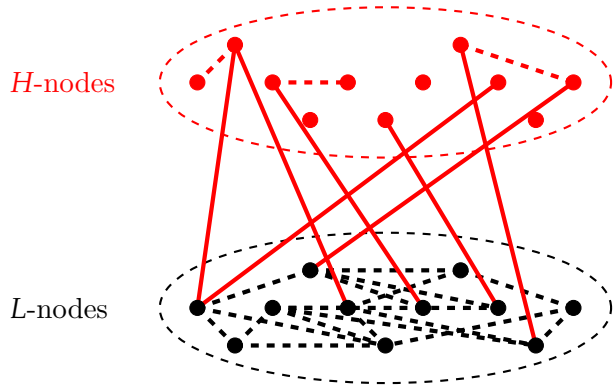
\begin{figure}[htbp]
 \begin{centering}
  \def\radius{2.6}
\def \Pointsize {1.4pt}
\begin{tikzpicture}[pre/.style={<-,shorten <=1.5pt,>=stealth,thick}, post/.style={->,shorten >=1pt,>=stealth,thick}]
\tikzstyle{every node}=[draw,shape=rectangle,minimum size=5mm, inner sep=0];
\tikzstyle{edge} = [draw,thick,-]
\tikzstyle{every node}=[shape=circle,minimum size=8mm, inner sep=0];

\draw[thick, dashed, red] (+3,3) ellipse (3 and 1);

\draw[thick, dashed] (+3,0) ellipse (3 and 1);

\node [right, red] at (-2,3) {$H$-nodes};
\node [right] at (-2,0) {$L$-nodes};


\draw[fill, red] (0.5,3) circle [radius=0.1];
\draw[fill, red] (1,3.5) circle [radius=0.1];
\draw[fill, red] (1.5,3) circle [radius=0.1];
\draw[fill, red] (2,2.5) circle [radius=0.1];
\draw[fill, red] (2.5,3) circle [radius=0.1];
\draw[fill, red] (3,2.5) circle [radius=0.1];
\draw[fill, red] (3.5,3) circle [radius=0.1];
\draw[fill, red] (4,3.5) circle [radius=0.1];
\draw[fill, red] (4.5,3) circle [radius=0.1];
\draw[fill, red] (5,2.5) circle [radius=0.1];
\draw[fill, red] (5.5,3) circle [radius=0.1];


\draw [dashed, ultra thick] (.5,3-3) -- (1, - 0.5);
\draw [dashed, ultra thick] (.5,3-3) -- (2,0.5);
\draw [dashed, ultra thick] (.5,3-3) -- (3,-0.5);
\draw [dashed, ultra thick] (1,2.5-3) -- (1.5,0.0);
\draw [dashed, ultra thick] (1,2.5-3) -- (3,-0.5);
\draw [dashed, ultra thick] (1.5,3-3) -- (2.5,0);
\draw [dashed, ultra thick] (1.5,3-3) -- (5,-0.5);
\draw [dashed, ultra thick] (1.5,3-3) -- (3,-0.5);
\draw [dashed, ultra thick] (1.5,3-3) -- (4.5,0);
\draw [dashed, ultra thick] (2,3.5-3) -- (3.5,0);
\draw [dashed, ultra thick] (2,3.5-3) -- (4,0.5);
\draw [dashed, ultra thick] (2,3.5-3) -- (4.5,0);
\draw [dashed, ultra thick] (2.5,3-3) -- (4,0.5);
\draw [dashed, ultra thick] (2.5,3-3) -- (5,-0.5);
\draw [dashed, ultra thick] (2.5,3-3) -- (3,-0.5);
\draw [dashed, ultra thick] (5.5,0) -- (4,0.5);
\draw [dashed, ultra thick] (5.5,0) -- (5,-0.5);
\draw [dashed, ultra thick] (5.5,0) -- (3,-0.5);

%

\draw [ultra thick, red] (.5,3-3) -- (4.5, 3);
\draw [ultra thick, red] (2,0.5) -- (5.5, 3.0);
\draw [ultra thick, red] (2.5,3-3) -- (1, 3.5);
\draw [ultra thick, red] (3.5,3-3) -- (1.5, 3);
\draw [ultra thick, red] (4.5,0) -- (3, 2.5);
\draw [ultra thick, red] (5,-0.5) -- (4, 3.5);
\draw [ultra thick, red] (0.5,0) -- (1, 3.5);

\draw [dashed, ultra thick, red] (4,3.5) -- (5.5, 3);
\draw [dashed, ultra thick, red] (1,3.5) -- (0.5, 3);
\draw [dashed, ultra thick, red] (1.5,3) -- (2.5, 3);


\draw[fill] (.5,3-3) circle [radius=0.1];
\draw[fill] (1,2.5-3) circle [radius=0.1];
\draw[fill] (1.5,3-3) circle [radius=0.1];
\draw[fill] (2,3.5-3) circle [radius=0.1];
\draw[fill] (2.5,3-3) circle [radius=0.1];
\draw[fill] (3,2.5-3) circle [radius=0.1];
\draw[fill] (3.5,3-3) circle [radius=0.1];
\draw[fill] (4,3.5-3) circle [radius=0.1];
\draw[fill] (4.5,3-3) circle [radius=0.1];
\draw[fill] (5,2.5-3) circle [radius=0.1];
\draw[fill] (5.5,3-3) circle [radius=0.1];

\end{tikzpicture}

  \end{centering}
  \caption{The typical graph in the heterogenous model $0 < \rho < 1$. Edge probabilities are $(c/n)^2$  in the H-H graph, $p^2$ in the L-L graph and $pc/n$ in the H-L graph.}
  \label{fig:pattern3}
\end{figure}

We analyze a  simple greedy algorithm termed {\it Chunk Matching} ($CM$) which finds allocations each time a given number of new pairs has joined the pool. Before we describe the chunk matching algorithm, we study the structure of the compatibility graph; the graph is composed of $3$ parts: (i) the H-H graph, which is the graph induced by the H pairs, (ii) the H-L graph which includes all nodes and only the edges between nodes of different types, and (iii) the L-L graph, which is  the graph induced by the L pairs (see Figure \ref{fig:pattern3}).

$CM$ receives as input two chunk sizes, $S_H$ and $S_L$, that determine the waiting times before making decisions. In particular, after the arrival of $S_H$ new nodes, it finds a maximum allocation in the graph composed of the union of the H-H and H-L graphs, ignoring edges between L nodes. After receiving $S_L/S_H$ chunks each of size $S_H$,  it  also finds a maximum allocation including the L-L  edges. If $S_H = S_L$, then both H and L types wait the same amount before being considered, but we still slightly favor the H-nodes (we first do matching in the graph without edges between L-L pairs and then we consider the entire graph), trying to compensate for the fact that they have fewer options and so are harder to match. Allowing $S_H < S_L$ does not only give priority to H-nodes, but also provides H nodes more matching opportunities by letting  L-nodes wait longer. We next formalize $CM$ as Algorithm \ref{alg:chunk}. At any time $t$, let $G_{t}$ be the residual graph with only unmatched pairs.



{Observe that $CM$  with $S_L = S_H = n$ can result in less allocations compared to the offline solution, as it prioritizes H nodes. However, in Lemma \ref{lem:two-stage} we prove that the difference in the number of allocations is $o(n)$. Thus, we often use $CM$  with $S_L = S_H = n$ as a proxy for the offline solution.}
Let  $M_C(S_H, S_L)$ denote the number of matches  obtained by $CM$ with chunk sizes of $S_H, S_L$. We will call the {\it online scenario} the case in which $S_L=S_H=1$.


{We analyze $CM$ for different regimes of waiting with H and L nodes. First, in Theorem \ref{thm:sublinear}, we show that waiting sublinearly with both H and L nodes will not increase the size of the matching significantly as compared to the case for which we do not wait with H nodes. Later, in Corollary \ref{cor:sublinear_vs_online}, we compare the sublinear waiting regime (with both H and L) to the online scenario (not waiting with neither H nor L), and show that the gain of sublinear waiting is small, i.e., $o(n)$.
On the other hand, waiting linearly with both H and L results in matching linearly more pairs (most of which are H nodes) as compared to the case for which we only wait with L nodes (Theorem \ref{thm:linear} Part \eqref{thm:linear:part2}). In Corollary \ref{cor:linear_vs_online}, we compare the linear waiting regime (with both H and L) to the online scenario and show that the gain of linear waiting is significant, i.e., $\Theta(n)$.
Furthermore, in Theorem \ref{thm:linear} part \eqref{thm:linear:part1}, we show that even if we divide the data into a ``few" chunks (or equivalently run the matching after $\beta n$ steps instead of waiting until the end) we will match linearly less nodes as compared to the offline solution. }

\begin{algorithm}[H]
\caption{Chunk Matching ($CM$)}
\label{alg:chunk}
\begin{algorithmic}[1]
\STATE Let $S_L$ be a divisor of $n$ and $S_H$ be a divisor of $S_L$; choose a maximum matching algorithm.
\item[{\bf For $\theta = S_L, 2S_L, \ldots, n$:}]
\item[{\bf For $\t = (\theta - S_L) + S_H, (\theta - S_L) +2S_H, \ldots, \theta$:}]
\STATE Run the maximum matching algorithm on the graph $G_{\t}$, ignoring edges between L nodes, breaking ties arbitrarily.
\STATE Remove the matched nodes.
\item[{\bf End for}]
\STATE Run the maximum matching algorithm on the graph $G_{\theta}$, breaking ties arbitrarily.
\STATE Remove the matched nodes.
\item[{\bf End for}]
\end{algorithmic}
\end{algorithm}



\begin{theorem}[Sublinear Waiting]
\label{thm:sublinear}
Suppose  $S = n^{1 - \epsilon}$ for some $0 < \epsilon < 1$ and $S$ is a divisor of $n$. Then
\begin{align*}
\E{M_C(S,S)} \leq \E{M_C(1,S)} + o(n).
\end{align*}
\end{theorem}

\begin{theorem}[Linear Waiting]
\label{thm:linear}
Let $0 < \beta < 1$ where $\beta n$ is a divisor of $n$.
\begin{enumerate}[(a)]
\item \textbf{Upper bound:}
There exists $\delta_{\beta} > 0$ such that:
\begin{align*}
\E{M_C(\beta n, \beta n)} \leq \E{M_C(n,n)} - \delta_{\beta} n.
\end{align*}
\label{thm:linear:part1}

\item \textbf{Lower bound:}
There exists $\delta'_{\beta} > 0$ such that:
\begin{align*}
\E{M_C(\beta n, \beta n)} \geq \E{M_C(1,\beta n)} + \delta'_{\beta} n.
\end{align*}
\label{thm:linear:part2}
\end{enumerate}
\end{theorem}
\noindent The next results show  that waiting only with  L pairs a constant fraction as opposed to not waiting at all will increase linearly the number of matched nodes (most of which are H nodes). Waiting with L pairs a ``sublinear time" however, will not increase the size of the matching significantly.
\begin{theorem}[Nonuniform Waiting] \
\label{thm:nonuniform}
\begin{enumerate}[(a)]
\item Let $0 < \gamma < 1$ where $\gamma n$ is a divisor of $n$.
 There exists $\delta_{\gamma} > 0$ such that:
    \label{thm:part1}
\begin{align*}
\E{M_C(1, \gamma n)} \geq \E{M_C(1,1)} + \delta_{\gamma} n.
\end{align*}

\item For any $S = n^{1 - \epsilon}$ where $0 \leq \epsilon < 1$:
\begin{align*}
\E{M_C(1,S)} \leq \E{M_C(1,1)} + O(S).
\end{align*}
\label{thm:part2}
\end{enumerate}
\end{theorem}


\noindent Theorem \ref{thm:sublinear} and part \eqref{thm:part2} of Theorem \ref{thm:nonuniform} imply that:
\begin{corollary}
\label{cor:sublinear_vs_online}
For any $S = n^{1 - \epsilon}$ in which  $0 < \epsilon < 1$ and $S$ is a divisor of $n$.
\begin{align*}
\E{M_C(S,S)} \leq \E{M_C(1,1)} + o(n).
\end{align*}
\end{corollary}

\noindent Also, part \eqref{thm:part2} of Theorem \ref{thm:linear} and part \eqref{thm:part1} of Theorem \ref{thm:nonuniform} imply that:
\begin{corollary}
\label{cor:linear_vs_online}
There exists $\delta''_{\beta} > 0$ such that:
\begin{align*}
\E{M_C(\beta n, \beta n)} \geq \E{M_C(1,1)} + \delta''_{\beta} n.
\end{align*}
\end{corollary}

\noindent Intuitively, the L pairs will not be difficult to match (as we will show, for online matching with only L pairs, there is almost no efficiency loss in comparison to the offline solution), and when the graph is sparse enough there are almost no short cycles in the  H-H graph. Understanding how $CM$ works on the H-L graph (the graph induced by all nodes and edges which connect only two different types) will be crucial for our proofs.
Thus in order to prove Theorems \ref{thm:sublinear} and \ref{thm:linear},  we first prove, in Section \ref{sec:prelim}, closely related results for general sparse homogenous graphs (See Propositions \ref{prop:sublinear}, \ref{prop:linear1}, and \ref{prop:linear2}). Then, in Section \ref{subsec:proof:linear}, we build upon the results of Section \ref{sec:prelim} and prove Theorems \ref{thm:sublinear} and \ref{thm:linear}. The proof of Theorem \ref{thm:nonuniform} is given in Appendix \ref{appendixB}.

{
Finally note that even though the online scenario has the worst performance, it still matches $\Theta(n)$ nodes; indeed it finds a maximal matching, and the size of a maximal matching is at last half of the maximum. Formally,
there exists $\delta > 0$ such that: $\E{M_C(1,1)} \geq  \delta n$.
}

%% file: GeneralSparseGraphs.tex
\section{Chunk matching in sparse homogeneous random graphs}
\label{sec:prelim}

Analyzing the  $CM$ algorithm on the H-L graph is an important ingredient to prove our main results. This is equivalent to studying the $CM$ algorithm on dynamic sparse non-directed bipartite graphs with uniform edge probability. Our results are stated for both generalized non-bipartite and bipartite sparse random graphs.

In a {\it dynamic homogeneous random graph}, each of $n$ nodes arrive sequentially  and there is a non-directed edge between  each arriving node and each existing node with a given probability. 
A dynamic homogeneous random graph is thus a non-directed special version of the dynamic kidney exchange graph. Also the offline graph, in which all nodes have arrived, is in this case simply an Erd\"{o}s-R\'{e}nyi random graph.

In the dynamic homogeneous random graph, the $CM$ algorithm uses a single chunk size, $S$, and we denote by  $M_C(S)$ the number of matches it finds.
As we will see the qualitative behavior of $CM$ in different waiting regimes for this homogeneous model is similar to the ones described in Section~\ref{sec:chunk}.
Propositions \ref{prop:sublinear}, \ref{prop:linear1}, and \ref{prop:linear2} below are the counterparts of Theorem \ref{thm:sublinear}, Theorem \ref{thm:linear} part \eqref{thm:linear:part1}, and Theorem \ref{thm:linear} part \eqref{thm:linear:part2}, respectively.


\begin{proposition}
\label{prop:sublinear}
Consider a dynamic homogeneous random graph with edge probability ${d}/n$. For any $0 < \epsilon < 1$, and any $S = n^{1 - \epsilon}$ that is a divisor of $n$,
\begin{align*}
\E{M_C(S)} \leq \E{M_C(1)} + o(n).
\end{align*}
\end{proposition}

\begin{corollary}
\label{cor:linear3}
Consider the  H-L dynamic graph with $0 <\rho \leq 1$. For any $0 < \epsilon < 1$, and any $S = n^{1 - \epsilon}$ that is a divisor of $n$, $\E{M_C(S)} \leq \E{M_C(1)} + o(n)$.
\end{corollary}

{\noindent We provide here a proof sketch for Proposition \ref{prop:sublinear} (the entire proof is given in Appendix \ref{appendixB})}.
\begin{proof}[\emph{\bf Proof sketch of Proposition \ref{prop:sublinear}}]


The intuition of the proof is as follows: after each chunk arrives, and after removing the matched nodes, the residual graph (before the next chunk arrives) has no remaining edges. So, suppose now that $S$ new nodes arrive and form edges. The resulting graph after these arrivals will contain at most $O(S) = o(n)$ edges and thus is extremely sparse. It consists of $O(S)$ connected components each of size $O(1)$; we show that with high probability, each of these components is a tree with depth one (See Figure \ref{fig:pattern2}). The maximum matching in a disconnected graph is the union of the maximum matching of each of its connected components. Thus without loss of generality, the online algorithm will also find the  maximum matching in each of these components separately. For instance consider the example of Figure \ref{fig:pattern2}; when $r_1$ arrives, it forms its three edges. Now since, w.h.p., nodes $c_{1}$, $c_{2}$, and $c_{3}$ will not have any other neighbors in this arriving chunk (the filled nodes in Figure \ref{fig:pattern2}), the decision of an online algorithm and of $CM$  would be the same.

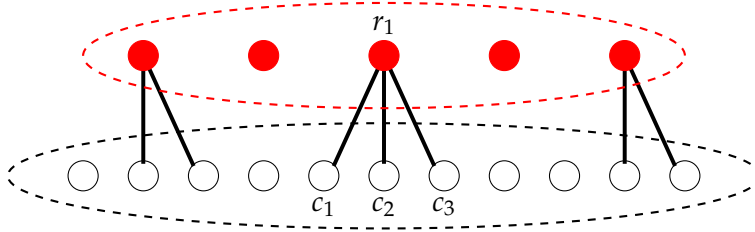
\begin{figure}[htbp]
 \begin{centering}
  \input{pattern2.tex}
  \end{centering}
  \caption{The typical connected components when the chunk is of a sublinear size; the filled nodes are the ones which arrived in the last chunk, and the not-filled nodes are those which arrived in the previous chunks, but which have not been matched yet.}
  \label{fig:pattern2}
\end{figure}
\end{proof}

\begin{proposition}
\label{prop:linear1}
Consider a dynamic homogeneous random graph with edge probability ${d}/n$. For any $0 < \beta < 1$ where $\beta n$ is a divisor of $n$, there exists $\delta_{\beta} > 0$ such that:
\begin{align*}
\E{M_C(\beta n)} \leq \E{M_C(n)} - \delta_{\beta} n.
\end{align*}
\end{proposition}
\begin{corollary}
\label{cor:linear1}
Consider the  H-L dynamic graph with $0 <\rho \leq 1$. For any $0 < \beta < 1$ where $\beta n$ is a divisor of $n$, there exists $\delta_{\beta} > 0$ such that $\E{M_C(\beta n)} \leq \E{M_C(n)} - \delta_{\beta} n$.
\end{corollary}

\noindent We provide here the main lines of the proof of Proposition \ref{prop:linear1}. 
We first need the following lemma about maximum matchings  (largest set of disjoint edges) in Erdos-Renyi random graphs. Let $G(n,\frac{{d}}{n})$ be  an undirected random graph $n$ nodes with edge probability $\frac{{d}}{n}$.
\begin{lemma}
\label{lem:sparse1}
The expected size of the maximum matching in $G(n,\frac{{d}}{n})$ is  $\alpha({d})n$, where $0 < \alpha(\cdot) <1$ is a strictly increasing function.
\end{lemma}
The proof can be derived from \citet{ks-revisit} (see also \citet{AshalgiGamarnikRoth}).

\begin{proof}[\emph{\bf Proof of Proposition \ref{prop:linear1}}]  Let $\mathcal{M}$ be a matching in a non-directed graph $G$. Note that $\mathcal{M}$ is not a maximum matching if it has an {\it augmenting path}, that is an odd length path $v_1,v_2,\ldots,v_{2l}$, where the even edges $(v_{2i},v_{2i+1})$ for all $i=1,\ldots,l-1$ are in $\mathcal{M}$ but the odd ones are not.

We  first prove the proposition for $S=\frac{n}{2}$, i.e., we show that by matching twice, once after $\frac{n}{2}$ nodes arrive and once after the last node arrives, $CM$ results in linearly many less matches than in the offline matching. To do this, we show there are linearly many disjoint augmenting paths for the union of the two matchings found by $CM$ with $S=\frac{n}{2}$.


By Lemma \ref{lem:sparse1}, the  expected size of the matching the $CM$ algorithm finds at time $n/2$ is $\alpha({d}) n/2$.
Denote by $Z_1$  the set of nodes arriving up to time  $n/2$ and are matched by  $CM$ at time $n/2$, and let $Z_2$ be the set of nodes that arrive after time $n/2$ and are not matched by the second matching. For any  $v_1,v_2 \in Z_1$ and $w_1, w_2 \in Z_2$, such that $v_1$ is matched to $v_2$ and the edges $(w_1,v_1)$ and $(v_2,w_2)$ exist, the path $p = (w_1,v_1)(v_1,v_2)(v_2,w_2)$ is an augmenting path. We call such augmenting paths {\it simple}. Denote by $P$ the set of simple augmenting paths.

In the following two claims, whose proofs are given in Appendix \ref{appendixB}, we state that the expected cardinality of  $P$ is  $\Theta(n)$, and that linearly many paths in $P$ are disjoint.

\begin{claim}
$\E{|P|} = \Theta(n)$.
\label{claim:sizeOfP}
\end{claim}
Intuitively a simple counting argument shows that $\E{|P|}$ scales as $\E{|Z_1| |Z_2|^2} \left({d}/n\right)^2$;  $|Z_2|$ is almost surely of linear size (note that many nodes in $Z_2$  are only connected to nodes in $Z_1$), and the expected size of $Z_1$ is linear in $n$ as well.


\begin{claim}
\label{claim:disjoint}
In expectation, the set $P$ consists of at least $\delta_{0.5} n$ vertex-disjoint paths.
\end{claim}

\noindent So far, we have considered the case  $S = n/2$. For $S = n/4$ similar arguments show that there exist $\delta_{0.25} >0$ such that $\E{M_C(n/4)} \leq \E{M_C(n/2)} - \delta_{0.25} n $, implying that $\E{M_C(S)}$ is strictly decreasing for
$S = 2^{-r} n$ for any positive constant $r$. Since this includes all possible divisors of $n$, we are done.\end{proof}

The next result is closely related  to Proposition \ref{prop:linear1}. It asserts that by setting the waiting periods to be  linear fractions of $n$, $CM$ results in linearly many more matches than the online matching.
\begin{proposition}
\label{prop:linear2}
Consider a dynamic random graph edge probability ${d}/n$. For any $0 < \beta < 1$ where $\beta n$ is a divisor of $n$  there exists $\delta'_{\beta} > 0$ such that:
\begin{align*}
\E{M_C(\beta n)} \geq \E{M_C(1)} + \delta'_{\beta} n.
\end{align*}
\end{proposition}
\begin{corollary}
\label{cor:linear2}
Consider the  H-L dynamic graph with $0 <\rho \leq 1$. For any $0 < \beta < 1$ where $\beta n$ is a divisor of $n$, there exists $\delta'_{\beta} > 0$ such that $\E{M_C(\beta n)} \geq \E{M_C(1)} + \delta'_{\beta} n$.
\end{corollary}
\begin{proof}[\emph{\bf Proof of Proposition \ref{prop:linear2}}]

We  first consider the first chunk of nodes, $\beta n$, and show that, after $\beta n$ nodes arrive, $CM$ matches linearly many more nodes than the online matching at that time.  Similarly to the proof of \ref{prop:linear1} we show that the residual graph of the online matching at time $\beta n$ contains linearly many disjoint augmenting paths.

Index the nodes by their arrival time, and denote by $\mathcal{M}$ the matching found by the online algorithm at time $\beta n$ and denote by $\Pi$ the set of augmenting paths in the graph at time $\beta n$ that have the following structure: there are four nodes, $i,i', j',j$ such that (a) $i$ is matched to $i'$ in $\mathcal{M}$ (or $\mathcal{M}(i)=i'$), (b) $j',\mathcal{M}(i)<i$, i.e., $j'$ and $\mathcal{M}(i)$ arrived before $i$, (c)  $j>i$,  i.e., $j$ arrived after $i$ and   (d) $j,j'$ are not matched in $\mathcal{M}$ (See Figure \ref{fig:pattern}).

Note that when the node $i$ arrives, the $CM$ online algorithm needs to decide whether to match it to $\mathcal{M}(i)$ and $j'$ (and maybe other existing nodes) and cannot predict that $\mathcal{M}(i)$ is the wrong choice in this case.

Note that the set of nodes that are not matched by $\mathcal{M}$ at time $\beta n$, denoted here by $Z$, is of size $\Theta(n)$, simply because online matches at most the same number of nodes as the maximum matching does, and by Lemma \ref{lem:sparse1}, we know that even the maximum matching leaves $\Theta(n)$ nodes unmatched.

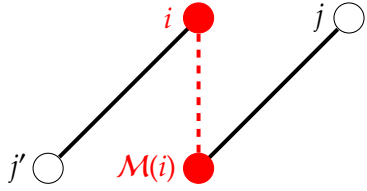
\begin{figure}[htbp]
 \begin{centering}
  \input{pattern.tex}
  \end{centering}
  \caption{An augmenting path in the set $\Pi$. The dashed edge belongs to matching $\mathcal{M}$ and the solid edges do not. $j'$ and $\mathcal{M}(i)$ arrived prior to $i$, and $j$ arrived after $i$.}
  \label{fig:pattern}
\end{figure}
The rest of the proof continues in a similar fashion to the proof of Proposition \ref{prop:linear1} with two claims whose proofs are given in Appendix \ref{appendixB}.
\begin{claim}
$\E{|\Pi|} = \Theta(n)$.
\label{claim:sizeOfPi}
\end{claim}
Similar to Claim \ref{claim:disjoint}, we can show that there exists $ \delta_1 >0 $ such that at least $\delta_1 n$ of these paths are disjoint, and can be added to $\mathcal{M}$ to construct a larger matching completing the proof for the first chunk of $\beta n$ nodes. The following claim asserts a similar result for the later chunks.
\begin{claim}
For any chunk $1 \leq {l} \leq n/S$, there exists $\delta_{l} > 0$ such that, in expectation, $CM$ matches $\delta_{l} n$ more nodes after chunk $c$ arrives.
\label{claim:laterChunks}
\end{claim}
Let $\delta'_{\beta} = \sum_{{l} = 1}^{n/S} \delta_{l}$; the above claim implies that at the end, the expected number of allocations of $CM$ is $\delta'_{\beta}n$ more than that of the online scenario, and this completes the proof of Proposition \ref{prop:linear2}.

\end{proof}


\section{Chunk matching on heterogeneous graphs and proofs of Theorems \ref{thm:sublinear} and \ref{thm:linear}}
\label{subsec:proof:linear}

The results in the previous section are given for the H-L graph only. In order to prove Theorems \ref{thm:sublinear} and \ref{thm:linear}, we still need to analyze $CM$ in the entire graph. Recall that the $CM$ algorithm gives priority to H pairs. In order to prove part  (\ref{thm:linear:part1}) of Theorem \ref{thm:linear}, the next lemma will provide the connection between the H-L graph and the entire graph, showing that we can essentially focus on the H-L graph. Before we state the lemma we consider the following procedure:

\vspace{5mm}
\hspace{25mm}
\begin{tabular}{| l |}
  \hline
\textbf{Two-Stage Matching:} \\
  \hline
a) Find the maximum matching in the H-L graph.\\
b) Find the maximum matching in the residual L-L graph.\\
  \hline
\end{tabular}
\vspace{5mm}

The next  lemma, whose proof is given in Appendix \ref{appendixB}, states that the matching obtained by the above procedure is nearly optimal:
\begin{lemma}
\label{lem:two-stage}
Let $\widetilde{\mathcal{M}}$ be matching obtained by the Two-Stage Matching procedure and $\mathcal{M}$ be the maximum matching; for $p_H = c/n$ and $p_L = p$,
\begin{align*}
\E{|\widetilde{M}|} \geq \E{|M|} - o(n).
\end{align*}
\end{lemma}
\begin{corollary}
\label{cor:two_stage}
A similar result holds for any chunk of the data with size $S \geq n^{1 - \epsilon}$ for $0 \leq \epsilon < 1$, i.e, the matching of the two-stage scheme produces is at most an $o(S)$ factor away from the one that a maximum matching algorithm will produce.
\end{corollary}
The L-L graph is a dense random graph. The following well-known theorem by Erd\"{o}s and R\'{e}nyi asserts that such dense graph, with high probability, has a perfect matching.
\begin{theorem}[Erd\"{o}s-R\'{e}nyi Theorem]
\label{thm:ER}
With high probability, an Erd\"{o}s-R\'{e}nyi random graph $G(\nu, \xi)$, with $\xi \geq \frac{\log{n}}{n}(1 + y)$ where $y >0$, has a perfect matching.
\end{theorem}

Corollary \ref{cor:two_stage} along with the Erd\"{o}s-R\'{e}nyi Theorem imply that when comparing the chunk matching with different chunk sizes, we can focus only on the H-L graph, since the remaining L nodes can always match to each other, and this does not result in a significant (if any) decrease in the number of allocations.


Theorem \ref{thm:sublinear} now follows from \ref{cor:linear3}, and Theorem \ref{thm:linear} now follows from Corollaries \ref{cor:linear1} and \ref{cor:linear2}. 

%% file: pattern2.tex
%
%
%
%

\def\radius{1.6}
\def \Pointsize {1.4pt}
\begin{tikzpicture}[pre/.style={<-,shorten <=1.5pt,>=stealth,thick}, post/.style={->,shorten >=1pt,>=stealth,thick}]
\tikzstyle{every node}=[draw,shape=rectangle,minimum size=5mm, inner sep=0];
\tikzstyle{edge} = [draw,thick,-]
\tikzstyle{every node}=[shape=circle,minimum size=8mm, inner sep=0];


\draw[line width=1.5pt] (0,+\radius) -- (0,0+0.16);
\draw[line width=1.5pt] (0,+\radius) -- (+.5*\radius-0.13, 0+0.13);
\draw[line width=1.5pt] (0,+\radius) -- (-.5*\radius+0.13, 0+0.13);
\draw[line width=1.5pt] (+2*\radius,+\radius) -- (+2.5*\radius-0.13, 0+0.13);
\draw[line width=1.5pt] (+2*\radius,+\radius) -- (+2*\radius, 0+0.16);
\draw[line width=1.5pt] (-2*\radius,+\radius) -- (-1.5*\radius-0.13, 0+0.13);
\draw[line width=1.5pt] (-2*\radius,+\radius) -- (-2*\radius, 0+0.16);

\draw [fill, red](0,+\radius) circle [radius=0.2];
\draw [fill, red](+\radius,+\radius) circle [radius=0.2];
\draw [fill, red](+2*\radius,+\radius) circle [radius=0.2];
\draw [fill, red] (-\radius,+\radius) circle [radius=0.2];
\draw [fill, red] (-2*\radius,+\radius) circle [radius=0.2];

\node [above] at (0,+\radius) {$r_1$};
\node [below] at (0,0) {$c_2$};
\node [below] at (+.5*\radius,0) {$c_3$};
\node [below] at (-.5*\radius,0) {$c_1$};

\draw (0,0) circle [radius=0.2];
\draw (+.5*\radius, 0) circle [radius=0.2];
\draw (-.5*\radius, 0)  circle [radius=0.2];
\draw (+\radius, 0) circle [radius=0.2];
\draw (-\radius, 0) circle [radius=0.2];
\draw (+1.5*\radius, 0) circle [radius=0.2];
\draw (-1.5*\radius, 0) circle [radius=0.2];
\draw (+2*\radius, 0) circle [radius=0.2];
\draw (-2*\radius, 0) circle [radius=0.2];
\draw (+2.5*\radius, 0) circle [radius=0.2];
\draw (-2.5*\radius, 0) circle [radius=0.2];




\draw[thick, dashed] (0,0) ellipse (5 and 0.7);
\draw[thick, dashed, red] (0,+\radius) ellipse (4 and 0.7);

\end{tikzpicture}

%% file: pattern.tex
%
%
%
%

\def\radius{2.6}
\def \Pointsize {1.4pt}
\begin{tikzpicture}[pre/.style={<-,shorten <=1.5pt,>=stealth,thick}, post/.style={->,shorten >=1pt,>=stealth,thick}]
\tikzstyle{every node}=[draw,shape=rectangle,minimum size=5mm, inner sep=0];
\tikzstyle{edge} = [draw,thick,-]
\tikzstyle{every node}=[shape=circle,minimum size=8mm, inner sep=0];

\draw[dashed, line width=1.5pt, red] (0,2) -- (0,0);
\draw[line width=1.5pt] (0,2) -- (-2+0.15,0.15);
\draw[line width=1.5pt] (0,0) -- (2-0.15,2-0.15);

\draw [fill, red](0,2) circle [radius=0.2];
\draw (2,2) circle [radius=0.2];
\draw [fill, red](0,0) circle [radius=0.2];
\draw (-2,0) circle [radius=0.2];

\node [left, red] at (0,2) {$i$};
\node [left] at (2,2) {$j$};
\node [left, red] at (-0.25,0) {$\mathcal{M}(i)$};
\node [left] at (-2,0) {$j'$};

\end{tikzpicture}

%% file: chainPcycle.tex
\section{Dynamic matching with short cycles and chains}
\label{sec:ChainPCycle}

\input{threeWay}
\input{chains}

%% file: threeWay.tex
\subsection{Waiting with $3$-ways}
\label{subsec:3Way}
Cycles of size 3 have been shown to increase efficiency in the static pools (\citet{RothKidneyAER}, \citet{AshlagiRothAER}). {Here, we generalize some of the results for the dynamic pools when 3-way cycles are also allowed. We slightly modify $CM$ to a chunk matching scheme denoted by  ${CM}^3$. The algorithm $CM^{3}$ also receives two chunk sizes as input; $S_H$ and $S_L$ where $S_H \leq S_L$. After each $S_H$ steps, $CM^{3}$ attempts to find a maximum allocation allowing cycles of length both 2 and 3 in the current compatibility graph excluding the L-L edges; after each $S_L$ steps, it finds the maximum number of exchanges in the whole remaining graph (including the L-L edges). Similar to the 2-way chunk matching, $CM^{3}$ also gives higher priority to matching H nodes by searching first for allocations excluding the L-L edges. Further, if $S_H < S_L$,  we make the L pairs (that can be easily matched fast) wait to help matching more H pairs.}

{
For the sake of brevity, we  focus on waiting at most a sublinear  time (i.e, chunk size $S = n^{1 - \epsilon}$ for some $0 < \epsilon < 1$). Proposition \ref{prop:sublinear:3way} extends Theorem \ref{thm:sublinear} for allowing also 3-way cycles, and its qualitative implication is the same: given that we wait with  L  nodes,  the difference (in the average number of matches) between waiting or not waiting with H nodes is not significant (more precisely, it is not linear in $n$).}

{
Interestingly, as we show in Theorem \ref{prop:sublinear:3way:LWaiting}, waiting  with L nodes even a sublinear time proves to be very effective in some regimes, and results in matching linearly more nodes compared to not waiting with L nodes at all.}

{Similar to $CM$, denote by ${M}^{3}(S_H, S_L)$ the number of matches that $CM^3$ with chunk sizes $S_H$ and $S_L$ finds. The following proposition and theorem are the main results of this subsection.}

{
\begin{proposition}[Sublinear waiting with H pairs]
\label{prop:sublinear:3way}
Suppose  $S = n^{1 - \epsilon}$ for some $0 < \epsilon < 1$ and $S$ is a divisor of $n$. Then
\begin{align*}
\E{M^{3}_C(S,S)} \leq \E{M^{3}_C(1,S)} + o(n).
\end{align*}
\end{proposition}}

{
The main difference in the analysis of $CM^3$ is that here we have a directed graph, and the residual graph does not consist of only isolated nodes anymore; it can contain many directed paths and even cycles of length greater than $3$. However, the residual graph mainly contains H nodes; similar to the proof of Lemma \ref{lem:greedy_dense}, we can show that at the beginning of each chunk the expected number of L nodes in the residual graph is $O(1)$.
Thus to compare the number of allocations of the  two schemes it suffices to compare the number of H nodes  they  match. Similar to $CM$ (with $S_H = 1$ and $S_L = S$), if we exclude the L-L edges, the graph  after a new chunk arrives is very disconnected; thus the decisions of $CM^3$ and and the online scheme result in almost the same number of matchings}. A formal proof is given in Appendix \ref{appendixC}.

{
\begin{theorem}[Sublinear waiting with L pairs]
\label{prop:sublinear:3way:LWaiting}
Let  $S = n^{1 - \epsilon}$ for some $0 < \epsilon < 1$ where $S$ is a divisor of $n$. If the parameters $\rho$, $p$, and $c$ satisfy the following condition:
\begin{align}
\label{eq:condition}
(1-p)(1- \rho)c e^{-c(1 + 2\rho)} - p \left(1 - e^{-c \rho}\right)\left(1 - c (1-\rho) e^{-c} - e^{-c(1-\rho)}\right) \geq \delta,
\end{align}
where $\delta > 0$ is a constant, then:
\begin{align*}
\E{M^{3}_C(1,S)} \geq \E{M^{3}_C(1,1)} + \delta n.
\end{align*}
\end{theorem}}

\begin{figure}[tbh]
\centering
{\includegraphics[width=0.8\textwidth,natwidth=200,natheight=320]{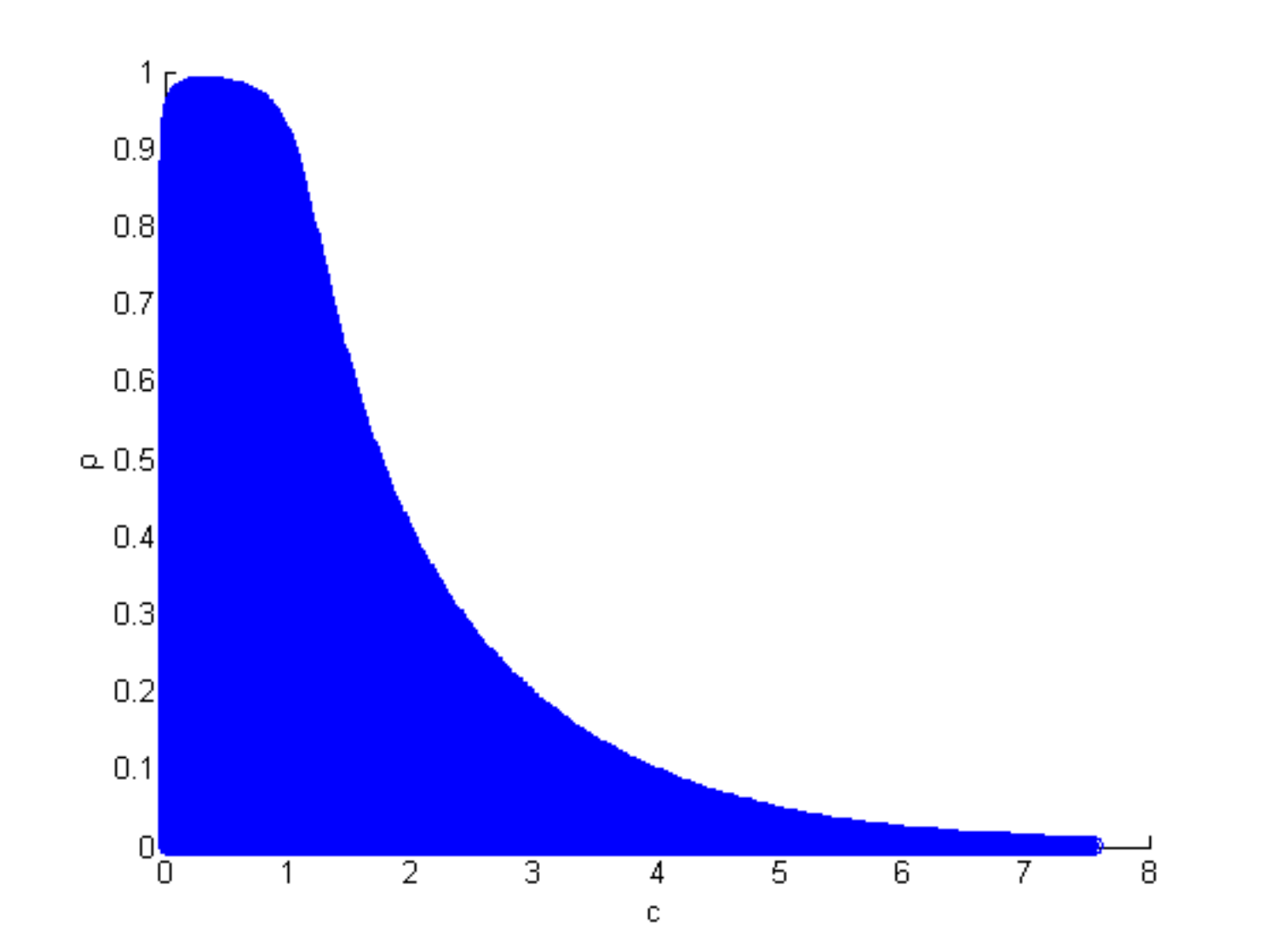}}
\caption{{\protect \footnotesize {The $(c,\rho)$ region satisfying Condition \eqref{eq:condition} for $p = 0.1$ and $\delta = 0.001$.}}}
\label{fig:3way:region}
\end{figure}

{The  proof for Theorem \ref{prop:sublinear:3way:LWaiting} is given
in Appendix \ref{appendixC}. The intuition for the result is as follows: In the online scenario, in many occasions, there will be a directed edge from an (arriving) L node $v$ to an (existing) H node $u$, but $u$ is not part of a cycle at that time. In fact,  there are (linearly) many such $v$ and $u$ nodes such that $v$ does not have a directed edge to any other H in the graph. Since $v$ is easy to match, the online scenario will ``quickly" find another cycle for the L node $v$ and the H node $u$ node will remain unmatched. However, under chunk matching, $v$ will have to wait and since it is an L node, it will be relatively ``easy" to close a 3-way cycle with $u$,$v$ and another L node arriving in the same chunk.  The proof deals with various subtleties such as ``harming" other H nodes by matching node $u$ too early. Figure \ref{fig:3way:region} shows a sample of the set of $(c, \rho)$ parameters that satisfies Condition \eqref{eq:condition} for $p = 0.1$ and $\delta = 0.001$.} 

{
The following corollary is a direct implication of Proposition \ref{prop:sublinear:3way} and Theorem \ref{prop:sublinear:3way:LWaiting}. Similar to corollary \ref{cor:sublinear_vs_online}, we compare the $CM^{3}$ with equal chunk sizes $S_H = S_L = S$ to the online scenario $S_H = S_L = 1$; however, here, for a certain set of parameters ($\rho$, $p$, and $c$), the gain of waiting for $S$ steps in 3-matching is linear in $n$ as opposed to the 2-matching where the gain is $o(n)$.}

{
\begin{corollary}
\label{cor:sublinear:3way:uniform}
Let  $S = n^{1 - \epsilon}$ for some $0 < \epsilon < 1$ where $S$ is a divisor of $n$. If parameters $\rho$, $p$, and $c$ satisfy Condition \eqref{eq:condition}, then,
there exits constant $\delta ' >0$ such that:
\begin{align*}
\E{M^{3}_C(S,S)} \geq \E{M^{3}_C(1,1)} + \delta' n.
\end{align*}
\end{corollary}}

%
%
%

%% file: chains.tex
\subsection{Dynamic matching with chains}
\label{subsec:chain}
In this section, we add an altruistic donor to the pool at time $t = 0$ and analyze how adding a single non-simultaneous  chain will affect the number of allocations. In particular, each time period a chain is found, the last node in the chain becomes a {\it bridge donor} (BD) for the next period.

Here, we only consider  the online scenario and analyze the following scheme:  after each new node arrives, we try to match it through a cycle of length at most $k$ or by adding it to the chain according to the following rules: (i) the bridge donor (last pair in the chain) must be of type H  and (ii) if the arriving node can form a $k$-way cycle with at least $k-1$ nodes of type H and can also form a path (of any length) connected to the BD, we break the tie in favor of the $k$-way. We denote such an online scheme by $\mathcal{O}^{c}_{k}$ and its counterpart without the chain by $\mathcal{O}_{k}$.

Observe that under $\mathcal{O}^{c}_{k}$ and $\mathcal{O}_{k}$ the residual graph does not contain any cycles with length smaller than or equal to $k$ (otherwise we would have performed such a cycle).
Our main results compare the performance of the greedy online matching with or without a chain ($\mathcal{O}^{c}_{k}$  and $\mathcal{O}_{k}$):

\begin{theorem}[Online Allocation with or without Chain] \label{thm:chain}
Consider the model given in \ref{subsec:model}; suppose we have one altruistic donor at time $0$;
\begin{enumerate}[(a)]
\item Suppose $\rho = 1$, i.e. all nodes are of type H, and $k = 3$; in expectation, $\mathcal{O}^{c}_{3}$ matches $\Theta(n)$ more nodes than $\mathcal{O}_{3}$.
    \label{thm:chain:Htype}
\item Suppose $0< \rho \leq 1$, but $k = 2$. In expectation, $\mathcal{O}^{c}_{2}$ matches $\Theta(n)$ more nodes than $\mathcal{O}_{2}$.
    \label{thm:chain:2way}
\end{enumerate}
\end{theorem}

The proof of Theorem~\ref{thm:chain} is presented in Appendix \ref{appendixC}. Interestingly, the proof with L nodes (part (b)) and without L nodes (part (a)) are very different. For example, in the first part without any L nodes the bridge donor is essentially ``forced" to wait for ``many" H nodes before it connects to one and by that time  a long path has been formed which allows the bridge donor to match many nodes at once. On the other hand, with L nodes, those long chains are not formed; when an  L node arrives it can form a cycle or a short chain relatively quickly preventing long paths with only H nodes to be formed.
However, with L nodes, we can construct a  solution with  many short  chain segments; the idea is to show that, after enough nodes have arrived, each time an L node arrives, with a constant probability,  the bridge donor will initiate a small chain segment by connecting to the newly arrived L node, and continuing to an isolated path containing only H nodes and of length at least 2; note that the H nodes in such a path have no other incoming edges, and thus can never be matched when allowing only 2-way cycles.


Part (b) of Theorem~\ref{thm:chain}  is the online version of the main result  of \citet{AshlagiAlgChains}, who show that in a static large sparse pool (equivalent to our offline solution)  chains add significantly to the number of matched pairs.

Note that Theorem \ref{thm:chain} does not cover the case in which we have both a mixture of H and L type nodes, and we also allow $3$-way cycles. We believe a similar comparison holds for this case as well, but we were not able to prove it. Thus we state it as the following conjecture.

\begin{conjecture}
Suppose $0< \rho \leq 1$ and $k = 3$; in expectation, $\mathcal{O}^{c}_{3}$ matches $\Theta(n)$ more nodes than $\mathcal{O}_{3}$ does.
\end{conjecture}

%
%

%% file: extension.tex
\section{Chunk matching in moderately dense graphs}
\label{sec:dense}

We have studied so far dense graphs (in Lemma \ref{lem:greedy_dense}) and sparse graphs where the probability for connecting to an H node is $\frac{c}{n}$. The sparsity level is a way to model  the graphs we observe in a given horizon. If kidney exchange grows, we expect to see ``denser" graphs in a given horizon. Therefore to complete the picture we study here graphs in less sparse regimes than $\frac{c}{n}$.


More precisely, in this section we assume that $p_H = c n^{-1+ \sigma}$ where $0 \leq \sigma < 1$ and show that most of the results proven in the previous sections also hold for  chunk sizes that scale with $1/p_H$. For the sake of brevity, we only provide the exact statement of these results for the homogeneous (non-directed) random graphs with edge probability ${c}n^{-1+ \sigma}$. Similar to the special case of $p_H=c/n$, these can be generalized to the heterogenous model with both H and L nodes.

The following proposition is the counterpart of Proposition \ref{prop:sublinear} and it states that if the chunk size is smaller than $n^{1-2 \sigma}$, then we will not gain significantly compared to the online scenario. The proof is given in the Appendix \ref{appendixC}.

\begin{proposition}
\label{prop:dense:sublinear}
Consider a dynamic random graph with edge probability ${c}n^{-1+ \sigma}$. For any $S = n^{1 - 2\sigma - \epsilon}$ (where $\epsilon >0$) that is a divisor of $n$,
\begin{align*}
\E{M_C(S)} \leq \E{M_C(1)} + o(n).
\end{align*}
\end{proposition}

\begin{remark}
\label{rem:dense2}
In a dynamic random graph with edge probability ${c}n^{-1+ \sigma}$ with $\sigma >0$, the gain we get by waiting for $n^{1 - 2\sigma} \leq S  \leq n^{1 - \sigma}$ steps in $CM$ is unknown. We conjecture that it is  $\Theta(n)$ as well.
\end{remark}

The next two propositions state that if we wait $\Theta(1/p_H)$, we loose a constant fraction of the matching compared to the case when we wait until the end. On the other hand, we gain a constant fraction compared to the case when we do not wait at all and match right away.

\begin{proposition}
\label{prop:dense:linear1}
Consider a dynamic random graph with edge probability ${c}n^{-1+ \sigma}$; for any $0 < \beta < 1$ where $\beta n^{1-\sigma}$ is a divisor of $n$, there exists $\delta_{\beta} > 0$ such that:
\begin{align*}
\E{M_C(\beta n^{1-\sigma})} \leq \E{M_C(n)} - \delta_{\beta} n.
\end{align*}
\end{proposition}

\begin{proposition}
\label{prop:dense:linear2}
Consider a dynamic random graph with edge probability ${c}n^{-1+ \sigma}$; for any $0 < \beta < 1$ where $\beta n^{1-\sigma}$ is a divisor of $n$  there exists $\delta'_{\beta} > 0$ such that:
\begin{align*}
\E{M_C(\beta n^{1-\sigma})} \geq \E{M_C(1)} + \delta'_{\beta} n.
\end{align*}
\end{proposition}

The proofs of Propositions \ref{prop:dense:linear1} and \ref{prop:dense:linear2} are identical to the proofs when $p = c/n$ and  are based on constructing augmenting paths.

\begin{remark}
\label{rem:dense1}
For any dynamic random graph with edge probability ${c}n^{-1+ \sigma}$ with $\sigma >0$, the Erd\"{o}s and R\'{e}nyi Theorem (Theorem \ref{thm:ER}) implies that, if we wait till all nodes arrive (or even wait $S = \Theta(n)$ steps) we will find a perfect matching with high probability.
\end{remark}

%% file: appendix.tex
\appendix

\section{Missing proofs of Section \ref{sec:model}}
\label{appendixA}

\begin{proof}[{\emph{\bf Proof of Lemma \ref{lem:greedy_dense}}}]
\label{app:lem:sgreedy_dense}
In order to prove Lemma \ref{lem:greedy_dense}, we study the process of the number of unmatched nodes at any time $t$; let $Z_t$ be the number of unmatched nodes at time $t$. We show that $\E{Z_n} = O(1)$. To do so, we use the basic property of any online greedy algorithm: if node $i$ and $j$ belong to $Z_t$ they could not be matched to each other, thus there is no edge between them (the probability of this event is $(1-p_H^2)$). The fact that there is no edge between any two of these $Z_t$ nodes gives us an upper bound on the probability that $Z_t$ is larger than one:

\begin{align*}
\P{Z_t = i} \leq (1 - p_H^2)^{i \choose 2} \leq  (1 - p_H^2)^{i-1}~~~ 2 \leq i \leq t.
\end{align*}

\noindent{Using this bound, we compute an upper bound for $\E{Z_t}$, where $t \geq 2$:}

\begin{align}
\label{lem:dense_exp}
\E{Z_t} &= \sum_{i = 1}^{t} \P{Z_t \geq i} \leq 1 +  \sum_{i = 2}^{t} \sum_{j = i}^{t} (1 - p_H^2)^{j-1} \nonumber \\
& \leq 1 + \frac{1 - p_H^2 -p_H^2(t-1)(1-p_H^2)^t}{p_H^4}
\end{align}

Since $p_H$ is here a constant independent of $n$, it follows that $\E{Z_n} = O(1)$. Finally note that $\E{|M_G|} = n/2 - 1/2\E{Z_n} = n/2 - o(n)$.
\end{proof}

%
%
%
%
%
%
%


\section{Missing proofs of Sections \ref{sec:chunk} - \ref{subsec:proof:linear}}
\label{appendixB}

\begin{proof}[\emph{\bf Proof of Proposition \ref{prop:sublinear}}]
We first prove the result for $S < n^{1/2}$, and then generalize it to the case where $n^{1/2} \leq S \leq n^{1 - \epsilon}$.

We begin by showing that the graph induced by the set of nodes in the arriving chunk $S$ contains no edges with high probability. Denote by
$\mathcal{E}$ the set of edges induced by the most recent chunk of nodes (the filled nodes in Figure \ref{fig:pattern2}).

\begin{align}
\E{|\mathcal{E}|} & = 1/2 \sum_{i,j \in S} \P{\tm{$i$ is connected to $j$}} \nonumber \\
& = \frac{|S|\left(|S| -1\right){d}}{2n} = O\left(\frac{S^2}{n}\right) = o(1).
\label{eq:lem:prf:exp_edges}
\end{align}


By Markov's inequality,

\begin{align*}
\P{|\mathcal{E}| > 1} \leq \E{|\mathcal{E}|} = o(1),
\end{align*}
implying that w.h.p. the set  $\mathcal{E}$ is empty.

Note that after a new chunk of $S$ nodes arrive, the graph consists of nodes from the previous residual graph (with no edges between themselves) and the new $S$ nodes. Next we show that after the arrival of the new $S$ nodes, w.h.p.  no node from the residual graph (not-filled nodes in Figure \ref{fig:pattern2}) has degree larger than one.
Denote by $\mathcal{C}$  the set of nodes of the residual graph. By union bound, we have:

\begin{align*}
\P{ \tm{$ \exists ~ i \in \mathcal{C}$ with degree more than 1}} & \leq \sum_{i \in \mathcal{C}} \P{\tm{$i$ has degree more than 1}} \\
&= |\mathcal{C}| \left[ 1 -  \P{\tm{$i$ has degree zero or one}}\right] \\
&= |\mathcal{C}| \left[ 1 - (1 - \frac{{d}}{n})^{|S|} - \frac{|S|{d}}{n} (1 - \frac{{d}}{n})^{|S|-1}\right]
\end{align*}
Using the well-known approximation that for small $x$, $(1 - x)^y = e^{-xy}\left( 1+ O(x^2y)\right)$, we have:

\begin{align}
\P{ \tm{$ \exists ~ i \in \mathcal{C}$ with degree more than 1}} & \leq |\mathcal{C}| \left( 1 - e^{-{d}|S|/n} - \frac{|S|{d}}{n} e^{-{d}|S|/n} + O(\frac{|S|}{n^2}) \right) \nonumber \\
&= O(\frac{|\mathcal{C}| |S|}{n^2}) = o(1),
\label{eq:lem:prf:deg}
\end{align}
where the last order equality holds because the size of $\mathcal{C}$ is at most $\Theta(n)$. We have $n/S$ chunks, and we showed that in each chunk, the gain of $CM$ over the online scenario is $o(S)$. Thus the total gain of $CM$ compared to online matching is $o(n)$.

Next we extend this analysis to the regime $n^{1/2} \leq S \leq n^{1 - \epsilon}$. The basic intuition is the same as for $S < n^{1/2}$; the subgraph of the arrived nodes is very sparse and w.h.p. there exists no nodes in the residual graph that has degree larger than one. The proof of the latter is the same as it is done for $S < n^{1/2}$; just consider \eqref{eq:lem:prf:deg}, the $\P{ \tm{$ \exists ~ i \in \mathcal{C}$ with degree more than 1}}$ is still $o(1)$ for $n^{1/2} \leq S \leq n^{1 - \epsilon}$. However, the proof of the former is different due to the fact that when we increase $S$ above $n^{1/2}$, the subgraph of the arrived nodes will have a few edges; in fact, Equation \eqref{eq:lem:prf:exp_edges} says that it has $O\left(\frac{S^2}{n}\right) = o(S)$ edges. Suppose we ignore these edges, then similar to the case $S < n^{1/2}$, we show that in each arriving chunk, $CM$ matches at most $o(S)$ more nodes that the online does. Now since adding $K$ edges to a graph increases the size of its maximum matching by at most $K$, it follows that when we add these $o(S)$ edges to the whole graph (both filled nodes and not-filled nodes), the size of its maximum matching increases by at most an $o(S)$ factor. Thus the gain of $CM$ over the online scenario in each chunk is $o(S)$, and we have $O(n/S)$ chunks, which implies that the overall gain of $CM$ with $S \leq n^{1 - \epsilon}$ is $o(n)$.
\end{proof}

\begin{proof}[\emph{\bf Proof of Claim \ref{claim:sizeOfP}}]

We need to prove that the cardinality of the set of simple paths is $\Theta(n)$. Note that
\begin{align*}
\E{|P|\big\vert Z_1, Z_2} = \frac{1}{4}|Z_1| |Z_2| \left(|Z_2| -1\right) \left(\frac{{d}}{n}  \right)^2,
\end{align*}
since $\frac{1}{4}|Z_1| |Z_2| (|Z_2| -1) $ counts the number of possible paths and the factor $\left(\frac{{d}}{n} \right)^2$ computes the probability of the two edges across the sets $Z_1$ and $Z_2$. Therefore

\begin{align}
\E{|P|} & = \frac{1}{4}\E{|Z_1| |Z_2| (|Z_2| -1)} \left(\frac{{d}}{n}  \right)^2.
\label{eq:Aug_path}
\end{align}

We will show that $Z_2$ is almost surely $\frac{|Z_2|}{0.5n} \geq e^{-{d}}$. This will prove the claim since

\begin{align*}
\E{|P|} & = \frac{1}{4}\E{|Z_1| |Z_2| (|Z_2| -1)} \left(\frac{{d}}{n}  \right)^2 \nonumber \\
& \geq \frac{1}{16}\alpha(c) n \left(e^{-c} n\right)^2 \left(\frac{{d}}{n}  \right)^2 + o(n) \nonumber \\
& = \zeta n + o(n),
\end{align*}

So, in the remainder of this proof,  let us show that $\frac{|Z_2|}{0.5n} \geq e^{-{d}}$ almost surely. Denote by $O$ the set of nodes with degree zero in the second chunk,  i.e, node $u$ that arrives after time $n/2$ belongs to $O$ if it has no edges. Clearly $|Z_2| \geq |O|$. We first show that $\E{|O|} \geq n/2 e^{-{d}}$. We then apply the Azuma's inequality to show that the $|O|$ is  concentrated around its expectation, and finally the Borel-Cantelli lemma to show that almost surely $|O| \rightarrow \E{|O|}$.

For each node $u$ that arrives after time $n/2$, the probability that it has degree zero is $(1 - \frac{{d}}{n})^{n/2 -1 + |Z_0|}$; recall that $Z_0$ is the set of nodes that arrived before time $n/2$ but did not get matched by the first allocation. For $n$ sufficiently large, $(1 - \frac{{d}}{n})^{n/2 -1 + |Z_0|} \geq e^{-{d}}$. Thus $\E{|O|} \geq e^{-{d}}n/2$. Next we show that the size of set $O$ is  concentrated around its mean. Note that $|O|$ is a function of all possible edges that may be formed at any time $1,\ldots, n$; denote this set by $\mathcal{E}$. For each such edge $e \in \mathcal{E}$, if it exists it may change the value of $|O|$ by at most $2$. Thus by the  Azuma's inequality to the corresponding Doob martingale (see \citet{Doob}), for any $\epsilon >0$:

\begin{align*}
\P{||O| - \E{|O||} \geq \epsilon n} & \leq 2 \left[\exp \left(- \frac{\epsilon^2n}{4 ({d}+ \epsilon')n}\right) \P{|\mathcal{E}|\leq ({d}/2+ \epsilon')n} \right.\\ &
+ \left.\exp \left(- \frac{\epsilon^2n}{4 {n \choose 2}}\right) \P{|\mathcal{E}| > ({d}/2+ \epsilon')n} \right]\\
& \leq 4 \exp\left( -\epsilon''n\right),
\end{align*}
where in the first line we condition on the size of the set $\mathcal{E}$ and use the fact that the number of edges formed (the size of $\mathcal{E}$) is  concentrated around its expectation (by Chernoff bounds) which is ${d}n/2 + o(n)$; further, $\epsilon'$ and $\epsilon''$ are positive constants used for the ease of presentation. Finally, note that:

\begin{align*}
\sum_{n = 1}^{\infty} \P{||O| - \E{|O||} \geq \epsilon n} < \infty,
\end{align*}
and the Borel-Cantelli lemma implies that $|O|/\E{|O|} \rightarrow 1$ almost surely.
\end{proof}

\begin{proof}[\emph{\bf Proof of Claim \ref{claim:disjoint}}]

We will find a lower bound on the number of disjoint paths. We  index the edges in the first matching arbitrarily by $(1,2), (3,4), \ldots, (|Z_1|-1, |Z_1|)$, and run the following iterative procedure: at iteration $i$, keep one path that contains edge $(2i-1,2i)$ (say path $p_i = (u_i, 2i-1,2i,v_i)$) and delete the others; also  delete all  paths that include either  $u_i$ or $v_i$. Clearly this procedure provides a set of disjoint paths. Next we compute the expected number of paths that remain after running this procedure. We compute the probability that we have at least one $p_i$ in the $i$-th iteration; suppose $i \leq 0.5 \min\{|Z_1|,|Z_2|\}$:

\begin{align*}
\P{\tm{at least one $p_i$}} = 2 {|Z_2| - 2i +2 \choose 2} \left(\frac{{d}}{n}\right)^2,
\end{align*}
on the other hand, if $i > 0.5|Z_2|$ then such probability will be zero. Summing over all $i \leq 0.5|Z_1|$, we have:

\begin{align}
\sum_{i = 1}^{0.5|Z_1|} \P{\tm{at least one $p_i$}}
& =  \sum_{i = 1}^{0.5 \min \{|Z_1|,|Z_2|\}} (|Z_2| - 2i +2) (|Z_2| - 2i +1) \left(\frac{{d}}{n}\right)^2 \nonumber \\
& \geq \left(\frac{{d}}{n}\right)^2 \sum_{i = 1}^{0.5 \min \{|Z_1|,|Z_2|\}} (|Z_2| - 2i )^2  \nonumber \\
& \geq 4 \left(\frac{{d}}{n}\right)^2 \sum_{j = 1}^{0.5 \min \{|Z_1|,|Z_2|\}} j^2.
\label{eq:disjoint}
\end{align}
 From the proof of the previous claim,  $\frac{|Z_2|}{0.5n} \geq e^{-{d}}$ almost surely. We will further show that  $\frac{|Z_1|}{0.5n} \geq {d}/2e^{-{d}}$ almost surely: recall that $Z_1$ is the set of nodes that were matched by the maximum matching at time $n/2$. Clearly the size of this matching is at least the number of isolated edges in the first chunk of the data. In expectation we have ${0.5n \choose 2} \frac{{d}}{n} \left[(1- \frac{{d}}{n})^{0.5n -1}\right]^{2}$ such edges. Similar to the proof of Claim \ref{claim:sizeOfP}, we can use the Azuma's inequality and the Borel-Cantelli lemma to prove that the number of isolated edges converges to its mean almost surely. This implies that $\frac{|Z_1|}{0.5n} \geq {d}/2e^{-{d}}$, and consequently that $\min \{|Z_1|,|Z_2|\} = 0.5 e^{-{d}} \min \{1,{d}/2\} n$ almost surely. Plugging this in \eqref{eq:disjoint} we obtain that

\begin{align*}
\sum_{i = 1}^{|Z_1|/2} \P{\tm{at least one $p_i$}} \geq 4 \left(\frac{{d}}{n}\right)^2 \frac{\left(0.25 e^{-{d}}  \min \{1,{d}/2\}n \right)^3}{3} = \delta_{0.5n} n,
\end{align*}
where $\delta_{0.5} = \frac{{d}^2 e^{-3{d}} \min \{1,{d}/2\}^3}{48}$.
\end{proof}

\begin{proof}[\emph{\bf Proof of Claim \ref{claim:laterChunks}}]
First note that we have already proved the claim for $l =1$. Next we prove for $l =2$;
the main difference between the first two chunks is that at time $\beta n$ there are some nodes left unmatched in both of these schemes. However, the size of these two residual graphs are not the same; before the new chunk arrives, the online matching has left the set $Z$ of nodes unmatched.
Similarly let $Z'$ be the set of unmatched nodes of the chunk matching with $S = \beta n$. We obtained that $|Z|\geq |Z'| + 2 \delta_1 n$ for some $\delta_1n$.


Note that both the  residual graphs are empty. Let us partition the set $Z$ into two subsets $Z_1$ with size $|Z'|$ and $Z_2$. Since all the nodes have degree zero and the future edge formations are independent and identical, the partition is arbitrary. We can look at the nodes in $Z_2$ as the nodes that were matched by the chunk matching and not by the online scenario. Thus in the second chunk, if the online scenario matches a node from the set $Z_2$, it only reduces its previous gap with the chunk matching. However, at time $2 \beta n$ we can compare the online and chunk matching on the sets $Z_1$ and the new arrived chunk similar to the way we compared them at the end of the first chunk, and get a similar result. Repeating similar arguments for the other chunks , and showing that in each chunk $l$, where $1 < l \leq \Delta$, there exists $\delta_l$ such that the online matching matches at least $2\delta_l n$ less nodes that the chunk with $S = \beta n$ does, completing the proof.

\end{proof}

\begin{proof}[\emph{\bf Proof of Lemma \ref{lem:two-stage}}]
\label{app:lem:tow-stage}


We will show that the number of disjoint augmenting paths in the symmetric difference of $\widetilde{M}$ and $M$ is at most $o(n)$.
Let  $\widetilde{M}_1$ ($\widetilde{M}_2$) be the matching obtained by running a maximum matching in the H-L (residual L-L) graph. Note that the H-L graph is a sparse bipartite Erd\"{o}s-R\'{e}nyi with edge probability $pc/n$. Similar to the proof of Lemma \ref{lem:sparse1}, we can show that
\begin{align*}
\E{|\widetilde{M}_1|} = 2\alpha'(pc/n) \min\{\rho n, (1 - \rho)n\},
\end{align*}
where $0 < \alpha'(\cdot) <1$ is a strictly increasing function. Thus after the first stage, there are $\Theta(n)$ L-nodes that are not matched. The graph induced by the remaining L nodes  is a random graph with edge  probability $p^2$ and thus by the Erdos-Renyi Theorem \ref{thm:ER} contains a perfect matching. This implies that after running the Two-Stage Matching, the residual graph contains no L-node. If there were no H-H edges, then the union of $\widetilde{M}_1$ and $\widetilde{M}_2$ would give us a maximum matching w.h.p. which implies that if we ignore the H-H edges, there is no augmenting path in the symmetric difference of $\widetilde{M}$ and $M$. Now let us add the edges between the H-nodes. Consider the symmetric difference of $\widetilde{M}$ and $M$: each augmenting path in this symmetric difference will include at least one H-H edge. Also note that the expected number of H-H edges is $c^2/2$, which is a constant. Thus the expected number of augmenting paths in the symmetric difference of $\widetilde{M}$ and $M$ is $o(n)$.
\end{proof}

\begin{proof}[\emph{\bf Proof of Theorem \ref{thm:nonuniform}}]

{We begin with the proof of part \eqref{thm:part1}. First note that Lemma \ref{lem:greedy_dense} and the Erd\"{o}s and R\'{e}nyi Theorem imply that in both schemes almost all the L nodes will be matched. Thus we focus on comparing the number of H nodes that these two schemes can match, and show that $CM$ with $S_L = \gamma n$ will match $\Theta(n)$ more H nodes.}

Let us focus on the first chunk and suppose we index the nodes by the time they arrive; suppose that at time $t < \gamma n/2$ and time $t+1$ two successive L nodes have arrived {and they are connected to each other}. Suppose node $t$  is not connected to any H node that has arrived before (this probability is at least $(1-pc/n)^{t-1}$). Similarly suppose node $t+1$  is not connected to any H node that has arrived before (again, this probability is at least $(1-pc/n)^{t-1}$). Now in the online matching, we would match node $t$ to an L node either at time $t$ or time $t+1$.

On the other hand, we show that in expectation we have $\Theta(1)$ nodes of type H that will arrive after $t$ in this chunk and are only connected to node $t$. Clearly if we wait until time $\gamma n$, we could have matched such an H node to node $t$. Thus by matching the L node $t$ along the way and not waiting until time $\gamma n$, we will decrease the size of matching by $\Theta(1)$ factor. {Summing over all $1 < t < \gamma n/2$, this implies that not waiting (online scenario) decreases the size of matching by $\Theta(n)$.}


{We can now start the detailed proof by introducing some notations.} Denote the event that nodes $t$ and $t+1$ are L nodes, connected to each other, and not connected to any available H nodes by $\mathcal{E}_t$. Further, let $\mathcal{H}_t$ be the set of H nodes that arrive after $t$ and are only connected to node $t$. First let us compute the probability of event $\mathcal{E}_t$:

\begin{align*}
\P{\mathcal{E}_t} \geq p^2(1 - \rho)^2 \left(1 - \frac{pc}{n}\right)^{2 \gamma n},
\end{align*}
next conditioned on event $\mathcal{E}_t$, let us count the set $\mathcal{H}_t$:
\begin{align*}
\E{|\mathcal{H}_t| \big \vert ~\mathcal{E}_t} & = \sum_{i = t+2}^{\gamma n} \P{\tm{$i$ is H node and only connected to $t$}~\big \vert ~\mathcal{E}_t} \\
&\geq \rho \left(\gamma n - t -2\right) \frac{pc}{n}\left(1 - \frac{pc}{n}\right)^{\gamma n -1}.
\end{align*}

\noindent{Putting these two together and summing over all $t < \gamma n/2$ gives us:}

\begin{align*}
\sum_{t = 1}^{\gamma n/2} \E{|\mathcal{H}_t| \I{\mathcal{E}_t}} & =
\sum_{t = 1}^{\gamma n/2} \E{|\mathcal{H}_t| \big \vert ~\mathcal{E}_t} \P{\mathcal{E}_t} \\
& \geq p^2(1-\rho)^2 (1 - \frac{pc}{n})^{3 \gamma n} \frac{pc\left(\gamma n/2\right)\left(\gamma n/2 -1\right)}{2n} = \kappa n + o(n).
\end{align*}
where $\I{\cdot}$ is the indicator function and $\kappa >0$ is just a constant used for the ease of presentation.

{So far, we have shown that at time $\gamma n$ (end of the first chunk), $CM$ with $S_L = \gamma n$ matches $\Theta(n)$ more nodes as compared to the online scheme. One can find similar patterns in later chunks (i.e., L nodes that could be matched to H nodes that arrive later in the same chunk, but will be matched to L nodes by the online scenario) as well and show that in any chunk, $CM$ with $S_L = \gamma n$ will match $\Theta(n)$ more nodes, and this completes the proof.}


We proceed to the proof of part \eqref{thm:part2}. Again, we focus on the first chunk and suppose we index the nodes by the time they arrive; in the online schemes (chunk with $S_H = S_L = 1$), at any time $1 \leq t \leq S_L$, if an L node arrives and it gets matched to an L node, it may cause a loss in the number of matching, because if the algorithm had waited before matching node $t$, this L node might have been used to match an H node that has arrived after time $t$. {However, we show that the probability of this event is ``small'': }We can have at most $S - t$ such H nodes, and each has an edge with node $t$ with probability $pc/n$. Thus by the union bound:
\begin{align*}
\P{\tm{L node $t$ could be used to match an H node}} \leq \frac{(S - t)pc}{n}.
\end{align*}

Using this upper bound, we compute an upper bound on the expected number of the mistakes that the online scheme can make:
\begin{align*}
\E{\tm{$\#$of mistakes of online scheme}} \leq \sum_{t = 1}^{S}\frac{(S - t)pc}{n} = \Theta\left(\frac{S^2}{n}\right).
\end{align*}

The same bound holds for any later chunk as well. We have $n/S$ chunks and for each of them the online makes at most $\Theta\left(\frac{S^2}{n}\right)$ mistakes, therefore the total number of mistakes is at most $O(S)$. This implies that chunk with $S_L = S$ and $S_H = 1$ matches at most $O(S)$ more than the online scenario ($S_H = S_L = 1$).
\end{proof}

\section{Missing proofs of Sections \ref{sec:ChainPCycle}-\ref{sec:dense}}
\label{appendixC}

\begin{proof}[\emph{\bf Proof of Proposition  \ref{prop:sublinear:3way}}]
Similar to the proof of Proposition \ref{prop:sublinear}, we show that after a new chunk arrives, the graph (the union of the residual graph and the new chunk {excluding the L-L edges}) is so disconnected that the decisions that the $CM^3$ with $S_H = S$ makes are mostly the same as those of $CM^3$ with $S_H = 1$ (See Figure \ref{fig:pattern3way}). {Further, in the second phase when searching  for an allocation in the entire graph (i.e., including the L-L edges), both of these schemes will ``see" almost the same residual graph, and thus find nearly the same number of exchanges.}

Consider the $k$-th chunk where $k = \Theta(n)$; the chunk consists of $\rho S + o(S)$ nodes of type H and $(1 - \rho) S + o(S)$ nodes of type L. The expected number of incoming edges to this new set of H nodes is $\Theta(S^2/n) = o(S)$, so w.l.o.g, we can ignore these edges. However, there are $\Theta(n)$ nodes of type H in the residual graph, and thus we have $\Theta(S)$ directed edges from the new chunk to the H nodes in the residual graph. With high probability, none of these outgoing edges will have the  same endpoint in the residual graph. More precisely, let $\mathcal{C}^3$ denote the set of H nodes in the residual graph. Similar to the calculation of \eqref{eq:lem:prf:deg}, one can show that the probability that there exists a node $i \in \mathcal{C}^3$ with more than one incoming edge from the new chunk is $o(1)$.

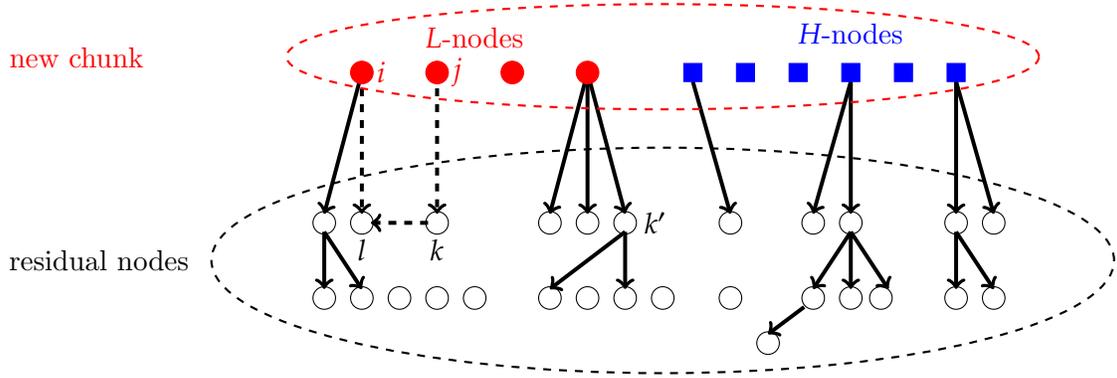
\begin{figure}[htbp]
 \begin{centering}
  \input{pattern3way.tex}
  \end{centering}
  \caption{The typical connected components when the chunk is of a sublinear size; the filled nodes are the ones arrived in the last chunk; the circle nodes are L-type and the square ones are H-types. The not-filled nodes arrived in the previous chunks, but have not been matched yet. The incoming edges to L-nodes are not shown.}
  \label{fig:pattern3way}
\end{figure}

Given the above observations, let us study the structure of the allocations after the new chunk arrives. We argue that the matching made by the $CM^3$ {(in the first phase when we ignore the L-L edges)} mainly consists of new L nodes and the nodes of set $\mathcal{C}^3$ (all the circle nodes  in Figure \ref{fig:pattern3way}); We showed that the number edges to new H nodes (and thus 2 and 3-way cycles involving them) is $o(S)$. Further, the number of 2 and 3-way cycles involving only H nodes is also $o(S)$ (in the entire graph there are $O(1)$ such cycles).   However, there are $\Theta(S)$ edges from the new L nodes to the residual H ones (i.e., the set $\mathcal{C}^3$). For each node in set $\mathcal{C}^3$, we know that it is unlikely that it has  more than one incoming edge from the new chunk. Fix a node  $v\in \mathcal{C}^3$.  We distinguish between two cases.

Suppose first that there doest not exist a node $w\in\mathcal{C}^3$ for which $v$ has an outgoing edge to such that $w$ also has an edge from a new L node (for example {$k'$} in Figure \ref{fig:pattern3way}), then if $CM^3$ chooses a cycle that contains $v$ it will also choose it in the online scenario.

Second, suppose $v$ has an outgoing neighbor $d \in \mathcal{C}^3$ such that  node $d$ also receives an edge from a new L node (for example, see nodes {$k$}, {$l$}, $i$, $j$, and the dashed edges in Figure \ref{fig:pattern3way}). This additional information may result in matching more nodes by the $CM^3$. For instance, in Figure \ref{fig:pattern3way}, if node $i$ arrives before node $j$ and edges $(l,i)$ and $(l,j)$ exist but edge $(k,i)$ does not (which happens with a constant probability), then the online will match {$l$} to $i$ and node {$k$} will remain unmatched, but the $CM^3$ will choose the $3$-way of $j-k-l$. Even though such mistakes are possible, we show that having these kind of structures is unlikely; For a connected pair of nodes $k, l \in \mathcal{C}^3$, the probability that both $k$ and $l$ receive edges from the new L nodes is ${S \choose 2} \left(c/n\right)^2$. We only have $\Theta(n)$ such connected pairs in the residual graph, thus the expected number of such patterns is $\Theta(S^2/n) = o(S)$. Thus the number of mistakes that the online scheme can make due to making early $2$-way cycles is $o(S)$. This shows that in a single chunk, $CM^{3}$ can only match at most $o(S)$ more nodes than the online does. Therefore, in the entire horizon, the gap between number of matches the two schemes achieve is of order $o(n)$.

\end{proof}

\begin{proof}[{\emph{\bf Proof of Theorem \ref{prop:sublinear:3way:LWaiting}}}]
\label{prof:prop:sublinear:3way:LWaiting}
{
Let $A^{CM}$ ($A^{O}$) be the set of H nodes that the $CM^{3}$ with $S_L = S$ (online, i.e., $CM^{3}$ with $S_L = 1$) matches in the entire horizon. We will show that when condition \eqref{eq:condition} holds then
$|A^{CM} \setminus A^{O}| \geq |A^{O} \setminus A^{CM}| + \delta n$ which implies the result,  because as mentioned before both schemes match almost all  L nodes. To do so, we find a lower bound on
$|A^{CM} \setminus A^{O}|$ and an upper bound on $|A^{O} \setminus A^{CM}|$. We find the lower bound by counting the number of some H nodes that $CM^3$ (with $S_L = S$) matches, but online can never match. On the other hand, we find the upper bound by counting the number of all H nodes that online may match, but $CM^3$ may not be able to match.}

\noindent {Consider the entire graph, and the set of H nodes $u$ with the following properties:
\begin{enumerate}
\item Node $u$ has only one incoming edge that is from an L node $v$.
\item Node $u$ has no outgoing edge to any H node.
\item Node $v$ arrives after $u$.
\item Node $v$ has no outgoing edge to any other H node.
\item Node $u$ does not form an edge to node $v$.
\end{enumerate}}
{
First let us evaluate the probability for having such H nodes $u$: the above five events are asymptotically independent, and (for large enough $n$) respectively have probabilities: $c (1-\rho) e^{-c}$, $e^{-c \rho}$, $1/2$, $e^{-c \rho}$, and $(1-p)$.\footnote{Note that for $n$ large enough, the indegree of H nodes has a Poisson distribution with rate $c$. Also, the outdegree of L/H nodes to H nodes has Poisson distribution with rate $c\rho$.} Thus conditioned on the event that a node is H, the probability that these five properties hold is $(1/2)(1-p)(1-\rho) c e^{-c(1 +2\rho)}$.
}

{
Now we claim that any H node that has the above properties will, w.h.p., be matched by $CM^3$ but not by the online scheme. Thus these H nodes will belong to the set $A^{CM} \setminus A^{O}$: note that node $u$ can only be matched in a 2-way or a 3-way that includes $v$. Because of the last property, $u$ and $v$ cannot form a 2-way, and the only way to match $u$ is by using 3-way cycles. Because node $u$ has no outgoing edge to another H node, it is not possible to form a H-H-L cycle with $u$; thus the only possible cycle is an H-L-L cycle. We show that $CM^{3}$ can easily form this cycle, but the online scheme cannot. Consider the chunk in which node $v$ arrives and forms the edge $(v,u)$. There exist $(1-\rho)S$ other L nodes in that chunk; each of these nodes can form a 3-way cycle with $u$ and $v$ with the constant probability $p^2$; thus, w.h.p., $CM^3$ can find such a 3-way at the end of the chunk. Also, note that in this chunk, node $v$ can be part of other cycles as well, but since it has no other outgoing H neighbor, those cycles can be either L-L or L-L-L cycles. Since we give priority to matching more H nodes, the H-L-L cycle including $v$ has the priority and $CM^3$ will choose it. On the other hand, w.h.p., the online scheme will match $v$  in some other cycle right after it arrives, and node $u$ will remain unmatched.
}

{
Now we compute an upper bound on $|A^{O} \setminus A^{CM}|$ using the H nodes $u$ that have the following properties:
\begin{enumerate}
\item Node $u$ has indegree at least $2$, and at least one of the incoming neighbors is an L node $v$.
\item Node $u$ has at least one outgoing edge to another H node $u'$.
\item Node $u'$ has an edge to node $v$.
\end{enumerate}}

{
Again, let us compute the probability that these three events happen: conditioned on $u$ being H, the probability of the intersection of the above events is $(1/2)p (1 - e^{-c \rho}) \left(1 - c (1- \rho) e^{-c} - e^{-c(1-\rho)}\right)$.
}

{
We show that any H node that has the above properties may be used by the online scheme to match {\em another H node} that $CM^3$ may not be able to match: suppose node $u$ is matched by $CM^3$ at time $t$, but not by the online. At some later time $t'$ in a later chunk, node $u$ will be in the residual graph of the online scheme (but not in the residual graph of $CM^3$). Suppose that node $u$ has an outgoing edge to another H node $u'$ and at time $t'$, node $u'$ has not been matched by neither online nor $CM^3$. Assume that at time $t'$ an L node arrives and it forms an outgoing edge to node $u$, and it has an incoming edge form $u'$; now the online scenario can form the H-H-L 3-way cycle $u-u'-v$, but because $CM^3$ has already matched $u$ in a previous chunk it cannot form this cycle. In the worst case, $CM^3$ will never be able to match $u'$ in any future chunk, and node $u'$ will belong to the set $A^{O} \setminus A^{CM}$. The above three properties are minimally required for having such $u-u'-v$ cycles, and thus this gives us an upper bound on $|A^{O} \setminus A^{CM}|$.
}

{
Given these two bounds, $|A^{CM} \setminus A^{O}| \geq (n/2) (1-p)(1-\rho) c e^{-c(1 +2\rho)}$ and $|A^{O} \setminus A^{CM}| \leq (n/2) p (1 - e^{-c \rho}) \left(1 - c (1- \rho) e^{-c} - e^{-c(1-\rho)}\right)$, condition \eqref{eq:condition} implies that $CM^3$ with $S_L = S$ matches at least $\delta n$ more nodes than $CM^3$ with $S_L = 1$ (i.e., the online scheme) does.
}
\end{proof}

\begin{proof}[\emph{\bf Proof of part \eqref{thm:chain:Htype} of Theorem \ref{thm:chain}}]
Observe that the offline graph contains only $O(1)$ cycles of length $2$ or $3$ since the expected number of cycles of length constant $k$ is ${n \choose k} \left(c/n\right)^k = O(1)$. Thus the size of matching obtained by the online scenario without a chain (i.e, $\mathcal{O}_{3}$) will be $O(1)$. On the other hand, we show that, in expectation, $\mathcal{O}^{c}_{3}$ can match $\Theta(n)$ nodes thus proving the claim: Consider the arriving process and suppose we index the nodes by the time they arrive. Let $i$ be the first arriving node that is connected to the altruistic donor. Clearly no chain has been formed before time $i$. Also suppose that no $2$- or $3$-way cycles have been performed either (this happens with a constant probability). Notice that $\E{i} = n/c$ and for any $\epsilon$, we have $\P{i>\epsilon n/c} = \left(1 - c/n\right)^{\epsilon n/c} = e^{-\epsilon} + o(1)$. Conditioning on the two events that $\{i>\epsilon n/c\}$ and no nodes were matched before time $i$, we compute a lowerbound on the expected size of the matching obtained by the online matching with a single chain; The residual graph at time $i$ is simply a directed Erd\"{o}s-R\'{e}nyi graph with at least $\epsilon n/c$ nodes and edge probability $c/n$. With high probability, such a graph has a path of length $\Theta(n)$ \citet{SudakovLongChains}. Let $p = (p_1, p_2, \ldots, p_L)$ be such a path. The probability that $i$ has an outgoing edge to at least one of the $p_j$'s for $1 \leq j \leq L/2$ is $1 - (1 - c/n)^{L/2}$. Now since $L = \Theta(n)$, this probability is bounded away from zero, implying that with a constant probability the chain formed at time $i$ matches at least $L/2 = \Theta(n)$ nodes. Finally, note that the expected number of total allocations is at least the number of nodes matched at time $i$; this completes the proof.
\end{proof}

\begin{proof}[\emph{\bf Proof of part \eqref{thm:chain:2way} of Theorem \ref{thm:chain}}]
As usual one can show (similar to the proof of Lemma \ref{lem:greedy_dense}) that at any time $t$ we have only $O(1)$ nodes of type L in the residual graph. Thus we know that both schemes will match almost all  L nodes, and hence it suffices to compare the number of H nodes matched by these two schemes. In particular, let $A^{c}_{t}$ ($A_{t}$) be the set of allocated H nodes by $\mathcal{O}^{c}_{2}$ ($\mathcal{O}_{2}$) by time $t$ (i.e, up to time $t$). We aim to show that

\begin{align}
\E{|A_{n}^c \setminus A_{n}|} \geq \E{|A_{n} \setminus A_{n}^c|} + \Theta(n).
\label{eq:chain}
\end{align}

To do so, we study the evolution of the two sets $A_{t}^c $ and $A_{t}$, and in particular their differences. The proof will follow from the next two claims that show that $\E{|A_{n} \setminus A_{n}^c|} = o(n)$ and that $\E{|A_{n}^c \setminus A_{n}|} = \Theta(n)$, implying together inequality \eqref{eq:chain}.

\begin{claim}
For every $t$, $\E{|A_{n} \setminus A_{n}^c|} = o(n)$.
\end{claim}
\begin{proof}
Let $I_{t}=\E{|A_{t} \setminus A_{t}^{c}| - |A_{t-1} \setminus A_{t-1}^{c}|}$ be the expected increment in the number of nodes that $\mathcal{O}_{2}$ matches at time $t$ but $\mathcal{O}^{c}_{2}$ doesn't. To prove the claim we show that $I_t=o(1)$ for any $t$.
Consider the node arriving at time $t$ and distinguish between the following cases.
\begin{enumerate}[(a)]
\item Node $t$ is an H node and it is connected to the bridge donor.
\label{enum:case:1}  Observe that the contribution of this case to the $I_{t}$ is at most $2 \rho c/n$: the probability of case \eqref{enum:case:1} is $\rho c/n$, and at any time $t$, the maximum number of H nodes that $\mathcal{O}_{2}$ can match (and thus can add to $A_{t} \setminus A_{t}^c$ at the worst case) is $2$.

\item Node $t$ is H type and it is not connected to the bridge donor. \label{enum:case:2}
In this case, it is possible that $\mathcal{O}_{2}$ matches node $t$ to a node $u$, but $\mathcal{O}^{c}_{2}$ cannot do so, because node $u$ was matched before. Thus node $t$ will be added to $A_{t} \setminus A_{t}^c$. However, we argue that the probability of this event is $o(1)$: If node $u$ is H type, the probability of a having a H-H two-way is $\left(c/n\right)^2$ and since we only have $O(n)$ nodes of H type in our pool, the probability that such a two-way exists is at most $O(1/n)$. Next suppose that node $u$ is of type L. In this case the probability of a having a H-L two-way is $pc/n$, but in expectation, we only have $O(1)$ such L nodes in the residual graph, thus the probability of this event is of order $o(1)$ as well. Therefore, the contribution of this case to $I_{t}$ is $o(1)$.

\item Node $t$ is an L node  and  is connected to the bridge donor. \label{enum:case:3}

Note that scheme $\mathcal{O}_{2}$ can match at most one H node, say node $u$. If node $u$ was  matched already by $\mathcal{O}^{c}_{2}$, then $I_{t}$ would be zero. On the other hand, if node $u$ was not matched by $\mathcal{O}^{c}_{2}$ before, then $\mathcal{O}^{c}_{2}$ can also perform the $2$-way cycle $(t,u)$, and since we give the priority to the $2$-way with one H node, $\mathcal{O}^{c}_{2}$ will perform such a cycle, and again $I_{t}$ will be zero.

\item Node $t$ is L type and it is not connected to the bridge donor. \label{enum:case:4}

The analysis of this case is very similar to the previous case and again we can show that $I_{t} = 0$.
\end{enumerate}
\end{proof}

\begin{claim}
 $\E{|A_{n}^c \setminus A_{n}|} = \Theta(n)$.
 \end{claim}
 \begin{proof}
 It is enough to show that the expected increment in each step $t=\Theta(n)$ is $\Theta(1)$. Fix a step $t$. With probability $(1-\rho)p$ the arriving node is an L node which is also connected to the  bridge donor.
We show that with a constant probability, $\mathcal{O}^{c}_{2}$ can add a path of length at least $4$ to the chain that contains at least $2$ nodes of type H that $\mathcal{O}_{2}$ can never match through any two-ways, thus these two nodes will surely belong to $A_{n}^c \setminus A_{n}$. Consider the entire pool (i.e., the graph that we obtain if we wait until time $n$ and make no allocations). A constant fraction of the H type pairs in this pool have indegree one: more precisely, the probability that a node of type H has indegree one is $\frac{c(n-1)}{n}\left(1 - c/n\right)^{n-2}$. Suppose node $u$ is such a node with the only incoming edge $(v,u)$. With a constant probability node $v$ is also of type H and has only one incoming edge $(w,v)$ where $w$ is of type L. At time $t$ large enough ($t = \Theta(n)$), we will have a linear number of such isolated $(v,u)$ edges in both of our residual pools (i.e., the residual graph of $\mathcal{O}^{c}_{2}$ and the one of $\mathcal{O}_{2}$). Suppose at time $t$, case \eqref{enum:case:3} happens, and there is an edge from $t$ to one of these isolated directed edges, say edge $(v,u)$, (i.e., edge $(t,v)$ exists) and node $t$ has no other outgoing edges to any of the H nodes in the pool. This happens with a constant probability bounded away from zero. Also suppose there is no edge from $v$ to $t$. Note that $\mathcal{O}_{2}$ can never match neither $u$ nor $v$. However, $\mathcal{O}^{c}_{2}$ can add the path $(BD,t,v,u,\ldots)$ to the chain. Thus we prove that each time a new L node arrives, and it is connected to the BD, with a constant probability we add two nodes to $A_{t}^{c} \setminus A_{t}$ that can never be removed from $A_{j}^{c} \setminus A_{j}$ for $j \geq t$.
\end{proof}
\end{proof}

\begin{proof}[\emph{\bf Proof of Proposition \ref{prop:dense:sublinear}}]
The  proof is very similar to the  proof of \ref{prop:sublinear}; We  use the observations that at any time the residual graph has no edges, and the graph formed after a new chunk of nodes arrived is extremely  disconnected. Using the notation defined in the proof of \ref{prop:sublinear}, we first upperbound the probability that there exists a node $c \in \mathcal{C}$ with degree more than one:

\begin{align*}
\P{ \tm{$ \exists ~ i \in \mathcal{C}$ with degree more than 1}} & \leq \sum_{c \in \mathcal{C}} \P{\tm{$i$ has degree more than 1}} \\
&= |\mathcal{C}| \left[ 1 -  \P{\tm{$i$ has degree zero or one}}\right] \\
&= |\mathcal{C}| \left[ 1 - (1 - p_H)^{|S|} - |S|p_H (1 - p_H)^{|S|-1}\right] \\
& \leq |\mathcal{C}| \left( 1 - e^{-|S|p_H} - |S|p_H e^{-|S|p_H} + O(|S|p_H^2) \right)\\
&= O(|\mathcal{C}| |S|p_H^2) = O(|\mathcal{C}| n^{-1-\epsilon}) = o(1).
\end{align*}

This shows that the edges between the newly arrived nodes and the nodes from the previous chunks most likely form many depth-one trees. Further, we show that there are only few edges in the subgraph of the newly arrived nodes: the expected number of edges among the new nodes is $O(S^2 p_H) = O(n^{1-3{\sigma} - 2\epsilon}) = o(S)$. Putting these two observations together in a similar manner (as we did in the proof of \ref{prop:sublinear}) , will prove the proposition.
\end{proof}

%% file: pattern3way.tex
%
%
%
%

\def\radius{1}
\def \Pointsize {1.4pt}
\begin{tikzpicture}[pre/.style={<-,shorten <=1.5pt,>=stealth,thick}, post/.style={->,shorten >=1pt,>=stealth,thick}]
\tikzstyle{every node}=[draw,shape=rectangle,minimum size=5mm, inner sep=0];
\tikzstyle{edge} = [draw,thick,-]
\tikzstyle{every node}=[shape=circle,minimum size=8mm, inner sep=0];



\draw[line width=1.5pt, ->] (2.5* \radius +1.4*\radius- 0.0,+\radius- 0.12) -- (2.5* \radius +1.4*\radius- 0.0,-\radius + 0.12);
\draw[line width=1.5pt, ->] (2.5* \radius +1.4*\radius- 0.0,+\radius- 0.12) -- (2.5* \radius +1.4*\radius- 0.0 + 0.5*\radius,-\radius+ 0.12);
\draw[line width=1.5pt, ->] (2.5* \radius + 0 ,+\radius- 0.12) -- (2.5* \radius + 0 ,-\radius+ 0.12);
\draw[line width=1.5pt, ->] (2.5* \radius + 0 ,+\radius- 0.12) -- (2.5* \radius - 0.5* \radius+ 0 ,-\radius+ 0.12);
\draw [line width=1.5pt, ->] (2.5* \radius -2.1*\radius ,+\radius - 0.12) -- (2.5* \radius -2.1*\radius + 0.5* \radius,-\radius+ 0.12);
\draw [line width=1.5pt, ->](-3 * \radius+2*\radius,+\radius) -- (-3 * \radius+2*\radius - 0.5*\radius,-\radius+ 0.12);
\draw [line width=1.5pt, ->](-3 * \radius+2*\radius,+\radius) -- (-3 * \radius+2*\radius + 0.5*\radius,-\radius+ 0.12);
\draw [line width=1.5pt, ->](-3 * \radius+2*\radius,+\radius) -- (-3 * \radius+2*\radius,-\radius+ 0.12);

\draw [line width=1.5pt, ->](-3 * \radius - \radius,+\radius) -- (-3 * \radius - \radius - 0.5*\radius,-\radius+ 0.12);

\draw [line width=1.5pt, ->,dashed](-3 * \radius - \radius,+\radius) -- (-3 * \radius - \radius ,-\radius+ 0.12);
\draw [line width=1.5pt, ->,dashed](-3 * \radius,+\radius) -- (-3 * \radius,-\radius+ 0.12);
\draw [line width=1.5pt, ->,dashed] (-3 * \radius - 0.12 ,-\radius) -- (-3 * \radius - \radius + 0.12,-\radius);

\draw [line width=1.5pt, ->] (2.5* \radius +1.4*\radius- 0.0,-\radius-0.12) -- (2.5* \radius +1.4*\radius + 0.5* \radius- 0.0,-2*\radius+ 0.12);
\draw [line width=1.5pt, ->] (2.5* \radius +1.4*\radius- 0.0,-\radius-0.12) -- (2.5* \radius +1.4*\radius ,-2*\radius+0.12);
\draw [line width=1.5pt, ->] (2.5* \radius + 0 ,-\radius -0.12) -- (2.5* \radius + 0 ,-2*\radius+0.12);
\draw [line width=1.5pt, ->] (2.5* \radius + 0 ,-\radius -0.12) -- (3* \radius + 0 ,-2*\radius+0.12);
\draw [line width=1.5pt, ->] (2.5* \radius + 0 ,-\radius -0.12) -- (2* \radius + 0 ,-2*\radius+0.12);
\draw [line width=1.5pt, ->] (2* \radius -0.12 ,-2*\radius -0.12) -- (2.5* \radius -2.1*\radius + 1* \radius,-2.6*\radius + 0.12);
\draw [line width=1.5pt, ->] (-4.5 * \radius ,-1*\radius - 0.12) -- (-4 * \radius ,-2*\radius + 0.12);
\draw [line width=1.5pt, ->] (-4.5 * \radius ,-1*\radius - 0.12) -- (-4.5 * \radius ,-2*\radius + 0.12);

\draw [line width=1.5pt, ->](-0.5 * \radius,-\radius - 0.12) -- (-0.5 * \radius,-2*\radius + 0.12);
\draw [line width=1.5pt, ->](-0.5 * \radius,-\radius - 0.12)  -- (-1.5 * \radius,-2*\radius + 0.12);

\draw (2.5* \radius +1.4*\radius- 0.0,-\radius) circle [radius=0.15];
\draw (2.5* \radius +1.4*\radius- 0.0 + 0.5*\radius,-\radius) circle [radius=0.15];
\draw (2.5* \radius + 0 ,-\radius) circle [radius=0.15];
\draw (2.5* \radius - 0.5* \radius+ 0 ,-\radius) circle [radius=0.15];
\draw (2.5* \radius -2.1*\radius + 0.5* \radius,-\radius)circle [radius=0.15];
\draw (-3 * \radius+2*\radius - 0.5*\radius,-\radius) circle [radius=0.15];
\draw (-3 * \radius+2*\radius + 0.5*\radius,-\radius) circle [radius=0.15];
\draw (-3 * \radius+2*\radius,-\radius) circle [radius=0.15];
\draw (-3 * \radius - \radius - 0.5*\radius,-\radius) circle [radius=0.15];
\draw (-3 * \radius - \radius ,-\radius) circle [radius=0.15];
\draw (-3 * \radius,-\radius) circle [radius=0.15];

\draw (2.5* \radius +1.4*\radius- 0.0,-2*\radius) circle [radius=0.15];
\draw (2.5* \radius +1.4*\radius- 0.0 + 0.5*\radius,-2*\radius) circle [radius=0.15];
\draw (2.5* \radius +1.4*\radius- 0.0 - 1*\radius,-2*\radius) circle [radius=0.15];
\draw (2.5* \radius + 0 ,-2*\radius) circle [radius=0.15];
\draw (2.5* \radius - 0.5* \radius+ 0 ,-2*\radius) circle [radius=0.15];
\draw (2.5* \radius -2.1*\radius + 1* \radius,-2.6*\radius)circle [radius=0.15];
\draw (2.5* \radius -2.1*\radius + 0.5* \radius,-2*\radius)circle [radius=0.15];
\draw (-3 * \radius+2*\radius - 0.5*\radius,-2*\radius) circle [radius=0.15];
\draw (-3 * \radius+2*\radius + 0.5*\radius,-2*\radius) circle [radius=0.15];
\draw (-3 * \radius+2*\radius,-2*\radius) circle [radius=0.15];
\draw (-3 * \radius - \radius - 0.5*\radius,-2*\radius) circle [radius=0.15];
\draw (-3 * \radius - \radius ,-2*\radius) circle [radius=0.15];
\draw (-3.5 * \radius ,-2*\radius) circle [radius=0.15];
\draw (-3 * \radius,-2*\radius) circle [radius=0.15];
\draw (-2.5 * \radius,-2*\radius) circle [radius=0.15];
\draw (+0 * \radius,-2*\radius) circle [radius=0.15];

\draw [fill, red](-3 * \radius + 0,+\radius) circle [radius=0.15];
\draw [fill, red](-3 * \radius+\radius,+\radius) circle [radius=0.15];
\draw [fill, red](-3 * \radius+2*\radius,+\radius) circle [radius=0.15];
\draw [fill, red] (-3 * \radius-\radius,+\radius) circle [radius=0.15];

\draw [fill, blue](2.5* \radius + 0 - 0.12,+\radius- 0.12) rectangle (2.5* \radius + 0 + 0.12,+\radius+0.12);
\draw [fill, blue](2.5* \radius +0.7*\radius- 0.12,+\radius- 0.12) rectangle (2.5* \radius +0.7*\radius + 0.12,+\radius + 0.12);
\draw [fill, blue](2.5* \radius +1.4*\radius- 0.12,+\radius- 0.12) rectangle (2.5* \radius +1.4*\radius + 0.12,+\radius + 0.12);
\draw [fill, blue] (2.5* \radius -0.7*\radius- 0.12,+\radius- 0.12) rectangle (2.5* \radius -0.7*\radius + 0.12,+\radius + 0.12);
\draw [fill, blue] (2.5* \radius -1.4*\radius- 0.12,+\radius- 0.12) rectangle (2.5* \radius -1.4*\radius + 0.12,+\radius + 0.12);
\draw [fill, blue] (2.5* \radius -2.1*\radius- 0.12,+\radius- 0.12) rectangle (2.5* \radius -2.1*\radius + 0.12,+\radius + 0.12);

\draw[thick, dashed] (0,-1.5*\radius) ellipse (6 and 1.5);
\draw[thick, dashed, red] (0,+1.2\radius) ellipse (5 and 0.7);


\node [right, red] at (-8.7,+1.2*\radius) {new chunk};
\node [right] at (-8.7,-1.5*\radius) {residual nodes};

\node [above, red] at (-2.5*\radius,+0.8*\radius) {$L$-nodes};
\node [above, blue] at (+2.5*\radius,+0.8*\radius) {$H$-nodes};

\node [below] at (-3 * \radius - \radius ,-\radius + 0.05) {{$l$}};
\node [below] at (-3 * \radius,-\radius + 0.05) {{$k$}};
\node [right] at (-0.5* \radius,-\radius) {{$k'$}};

\node [right, red] at (-3 * \radius - \radius -0.15 ,+\radius ) {$i$};
\node [right, red] at (-3 * \radius -0.15,+\radius ) {$j$};

\end{tikzpicture}